\documentclass[sigconf]{acmart}
\AtBeginDocument{%
  }
\usepackage{multirow}
\usepackage{graphicx}
\usepackage{subcaption}
\usepackage{float}
\usepackage{cuted}  
\usepackage{caption}    
\setcopyright{acmlicensed}
\copyrightyear{2025}
\acmYear{2025}
\acmDOI{XXXXXXX.XXXXXXX}
\acmConference[SLW]{SciSoc LLM Workshop}{August 6th, 2025}{Toronto}
\acmISBN{978-1-4503-XXXX-X/2018/06}

\setcopyright{none}
\renewcommand\footnotetextcopyrightpermission[1]{} 

\begin{document}

\title{Predicting Field Experiments with Large Language Models}


\author{
  Yaoyu Chen\textsuperscript{\rm 1}\quad
  Yuheng Hu\textsuperscript{\rm 1}\quad
  Yingda Lu\textsuperscript{\rm 1}\\   
  \textsuperscript{\rm 1}The Department of Information and Decision Sciences,  
  College of Business Administration, University of Illinois at Chicago\\
  \{ychen563, yuhenghu, yingdalu\}@uic.com
}


\begin{abstract}
Large language models (LLMs) have demonstrated unprecedented emergent capabilities, including content generation, translation, and simulation of human behavior. Field experiments, on the other hand, are widely employed in social studies to examine real-world human behavior through carefully designed manipulations and treatments. However, field experiments are known to be expensive and time consuming. Therefore, an interesting question is whether and how LLMs can be utilized for field experiments. In this paper, we propose and evaluate an automated LLM-based framework to predict the outcomes of a field experiment. Applying this framework to 276 experiments about a wide range of human behaviors drawn from renowned economics literature yields a prediction accuracy of 78\%. Moreover, we find that the distributions of the results are either bimodal or highly skewed. By investigating this abnormality further, we identify that field experiments related to complex social issues such as ethnicity, social norms, and ethical dilemmas can pose significant challenges to the prediction performance.
\end{abstract}




\maketitle

\section{Introduction}
Field experiments allow researchers to manipulate variables of interest in a real-world setting, such as human behaviors like ad clicking or donation giving, establishing causal relationships between interventions and outcomes. They typically begin by designing an intervention aligned with a specific research question, randomly assigning participants to treatment or control groups, and then measuring outcomes under natural conditions. The resulting data are collected and analyzed to evaluate the causal impact of the intervention~\citep{temp1}. It is adopted by a wide range of disciplines across academia and industry, such as finance~\citep{if5,if6}, marketing~\citep {if1,if2}, and organizational studies~\citep{if3,if4}. In recent years, online field experiments, also known as A/B testing, have revolutionized numerous online platform designs, advertising strategies, etc. While effective, field experiments are also known to be expensive and time consuming. For example, some experiments may take months or years to conduct~\citep{tempo1}. Moreover, recruiting high-quality participants for the experiments is challenging and costly~\citep{tempo2}, potentially affecting the experiment's outcomes.

In recent years, Large language models (LLMs) have demonstrated unprecedented emergent capabilities, including content generation, translation, and simulation of human behavior. For example, existing studies have explored the alignments between simulated data generated by LLMs and real data collected from human participants across various aspects, including human responses, traits~\citep{10}, moral standards~\citep{8}, preferences~\citep{19}, and emotions~\citep{11}. Therefore, it is becoming increasingly popular to leverage LLMs for simulating human behaviors~\citep{tempo3,tempo4}. For example, there are a handful of studies in which scholars have successfully instructed LLMs to replicate existing lab experiments across several disciplines, including psychology~\citep{1,auto1}, sociology~\citep{15,18,auto2}, and economics~\citep{1,12}. Their aim is to replicate existing lab experiments by treating LLMs as participants in lab experiments. 


However, directly applying these works to field experiment simulations in our context has several challenges. First, they mainly focused on lab experiment settings. But field experiments are inherently more challenging to conduct than laboratory experiments, due to diverse participant backgrounds, complex workflows, and multifaceted treatment designs. Second, most of the previous studies mainly relied on manual processes, and because of that, tested a handful of experiments with a limited scope of topics. While some recent studies tested on relatively large data, the selected experiments were limited to Likert-based psychological or social surveys~\citep{auto1,auto2}. In order to explore LLMs' capabilities, robustness, and generalizability in field experiment simulation, we need to test them on a much larger scale and broader range of experiments on different topics.



In this paper, we fill the literature gap by proposing an automated LLM framework to predict the outcome of a wide range of field experiments. Our framework has several major components, such as an information extraction module, a variant generation module, and a prediction module. Specifically, the information extraction module extracts key experiment settings, while the variant generation module generates false variants as distractors to confuse LLMs. Finally, the prediction modules leverage two prompt templates with a Chain-of-Thought design, prompting the LLM to predict the outcomes of a field experiment. 

We test this framework on 276 field experiments reported in premier academic journals from 2000 to 2024. Those experiments contain a total of 1261 conclusions with a wide range of topics such as labor-market discrimination, educational incentives, household finance behavior, and the impact of healthcare enrollment. Without any alignment techniques or fine-tuning, our framework achieves an average prediction accuracy of 78\%. We also test for data memorization effects by examining prediction results on recent experiments that appeared in 2024, which would be less likely to be included in the training data of LLMs. More interestingly, we also find that the prediction results are either bimodal or highly skewed. For example, our framework achieves nearly 100\% prediction accuracy on 71\% of conclusions while it completely fails to predict 18\% of the conclusions with close to 0\% accuracy. Further analyses reveal that our LLM-based framework has limitations in predicting experiments related to complex social issues such as ethnicity, social norms, and ethical dilemmas.




Our research makes several contributions:
\begin{itemize}
\item[1] We extend the current literature on LLMs' emergent capabilities by demonstrating that LLMs can simulate field experiments by predicting conclusions. While using LLMs to simulate human behavior is well studied in recent years, to the best of our knowledge, this is the first work in the literature to replicate large-scale field experiments that require a more complex environment setting and workflow design.

\item[2] Our proposed framework enables the prediction of field experiment conclusions in a fully automated, large-scale fashion. As a result, our framework is robust and can be generalized to a wider range of downstream applications in field experiments.

\item[3] This paper examines the prediction performance and reveals the limitations of LLMs in field experiment simulations.
\end{itemize}


\section{Literature Review}
Our study is related to the LLM literature on simulating human behavior, with a particular focus on experimental simulations by LLMs. Here, we highlight our contributions by comparing and contrasting our work with existing studies.

While past literature mainly focused on agent-based social simulation~\citep{if7,if8,if9}, there is an increasing trend to adopt LLMs as simulation tools. Existing studies have found that LLMs’ ability to simulate human behavior stems from their possession of human-like reasoning skills and their adaptivity to personas of diverse characters~\citep{22,23}. Upon those features of LLMs, Aher et al.~\cite{1} proposed the concept of "Turing Experiment", in which LLMs are profiled as synthetic participants of experiments with integrated prompt of experimental settings and demographic information. Similarly, Horton~\cite{12} demonstrated LLMs’ ability to simulate lab experiments and promoted it as a method of experimental pilot testing. The usage of LLMs in Horton~\cite{12}'s work is similar to Aher et al.~\cite{1}, consisting of two stages: prompting LLMs as synthetic experimental participants and collecting responses from conversations, reporting a few successful simulation cases of lab experiments. Leng \& Yuan~\cite{15} harnessed a three-phase procedure to complete the lab experimental simulation, which includes the initialization phase, interaction phase, and decision analysis phase. In the initialization phase, separate conversations of GPT-4 are prompted as vanilla experimental participants without specifying demographic information. Then, in the interaction stage, synthetic participants are prompted with the actions of other participants and asked to take actions and rationales according to the experiment design. Last, agents' actions and rationales are collected for analysis to conclude. Leveraging this procedure, the study simulates five existing lab experiments. Manning et al.~\cite{18} proposed an automated framework for various lab experimental simulations. Although the entire workflow is divided into seven steps, it is essentially similar to the manual procedures of the three papers above. However, the sole input is the backstory of the experiment. Based on the input information and powered by an LLM, the framework continues with subsequent steps: identifying dependent and independent variables, generating treatment values, profiling agents, organizing the interaction of agents, collecting data, and establishing causal relationships and conclusions. The paper reports the results of four social scenarios. Besides the aforementioned studies, which were tested only on a small scale of cases, there are extant papers that implemented large-scale testing on LLM-based experimental replication. Cui et al.~\cite{auto1} replicated 154 psychological experiments. Specifically, they profiled LLMs as either students or adults, retrieving Likert-like responses from LLMs, which is oversimplified compared to most lab and field experiments. By comparison, Ashokkumar et al.~\cite{auto2} extended such a massive replication to 70 more complex survey experiments, though their selected experiments were limited exclusively to US social surveys and did not include any field experiments in the primary tests.

In summary, while existing studies discussed above pioneered the application of Large Language Models (LLMs) to laboratory experiment simulations, several research gaps remain unaddressed. For instance, existing methods are not tested on field experiments that are inherently more complex than laboratory experiments, encompassing diverse participant backgrounds, more intricate workflows, and multifaceted treatment designs. Besides, despite extensive discussions on LLM bias~\citep{if10,ijk4,ijk5} and the fact that recognized biases in LLMs—such as gender~\citep{ijk7} and social norm bias~\citep{ijk6}—can compromise performance on downstream tasks~\citep{27}, only one of these studies mentioned the impact of certain topics on the replication successful rate of psychological surveys~\cite{auto1}. It is still unclear how the joint effect of topics and sentiments would undermine the fidelity of field experiment prediction. Another limitation is that these studies tested their simulation strategies on only a small number of experiments, restricting generalizability. To address these gaps, we evaluated our proposed framework at scale on field experiments from published papers, demonstrating not only that LLMs can accurately predict the outcomes of established experiments, but also clarifying boundary conditions under which they cannot provide reliable experimental predictions. Furthermore, current automated simulation framework~\citep{18,auto1,auto2} are incompatible with scenarios involving complex treatments involving human-object interactions and limit only to lab experiments with Likert response or survey. By contrast, our framework offers a broadly adaptable approach that supports experiments across a wide range of contexts.

\section{Data Collection and Filtering}

To explore LLMs' ability to predict field experiments in a large-scale fashion, we first need to collect existing field experiments. Inspired by prior studies that focus on using LLMs to simulate human behaviors in lab experiments from existing psychology literature, we also consider field experiments studied in premier journals in economics, such as The Review of Economic Studies, American Economic Review, Journal of Political Economy, and The Quarterly Journal of Economics. It is worth noting that we focus on field experiments in economics because they are generally a larger scale in terms of participant size and robust in terms of careful design. 

\begin{figure}[h]
  \centering
  \includegraphics[width=\linewidth]{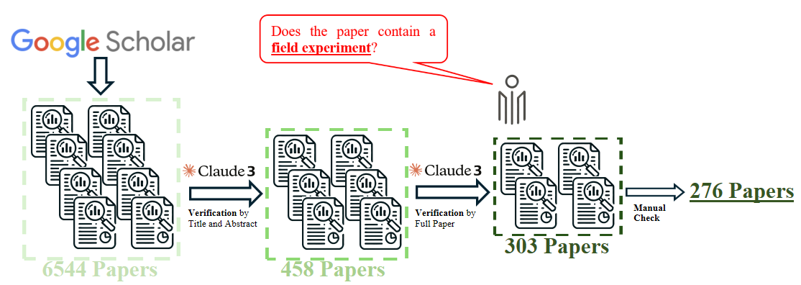}
  \caption{The Data Collection Workflow.}
  \Description{A flowchart illustrating the data collection process: from filtering 6544 papers to applying automated Claude-based checks and manual verification, resulting in 276 selected field experiments.}
  \label{fig:1}
\end{figure}

Figure~\ref{fig:1} shows the data collection and filtering workflow. Initially, 6544 papers containing keywords related to field experiments published between 2000 and 2024 of these top journals were selected. Then, we applied a two-layer verification process powered by Claude (Claude-3-opus-20240229). First, we prompted the title and abstract of each paper to Claude and asked it to judge if the paper designs and implements a field experiment. Upon this, the second verification prompted the entire paper to Claude, asking the same question. This strategy balances the accuracy of verification and the cost of calling Claude, as prompting the entire paper exponentially increases the cost. A final rule-based manual check to ensure the fidelity of the automated selection and filter out 276 papers for testing. More details about the manual check are available in Section~\ref{sec:Manual Robustness Check}. It is also worth mentioning that using Claude as the verification tool, instead of GPT, prevents potential data leakage as GPTs are leveraged as the prediction tool in our framework. The distribution of the selected papers is shown in Figure~\ref{fig:2} (In Appendix~\ref{app:data}).



\section{Framework}

\begin{figure}[h]
  \centering
  \includegraphics[width=\linewidth]{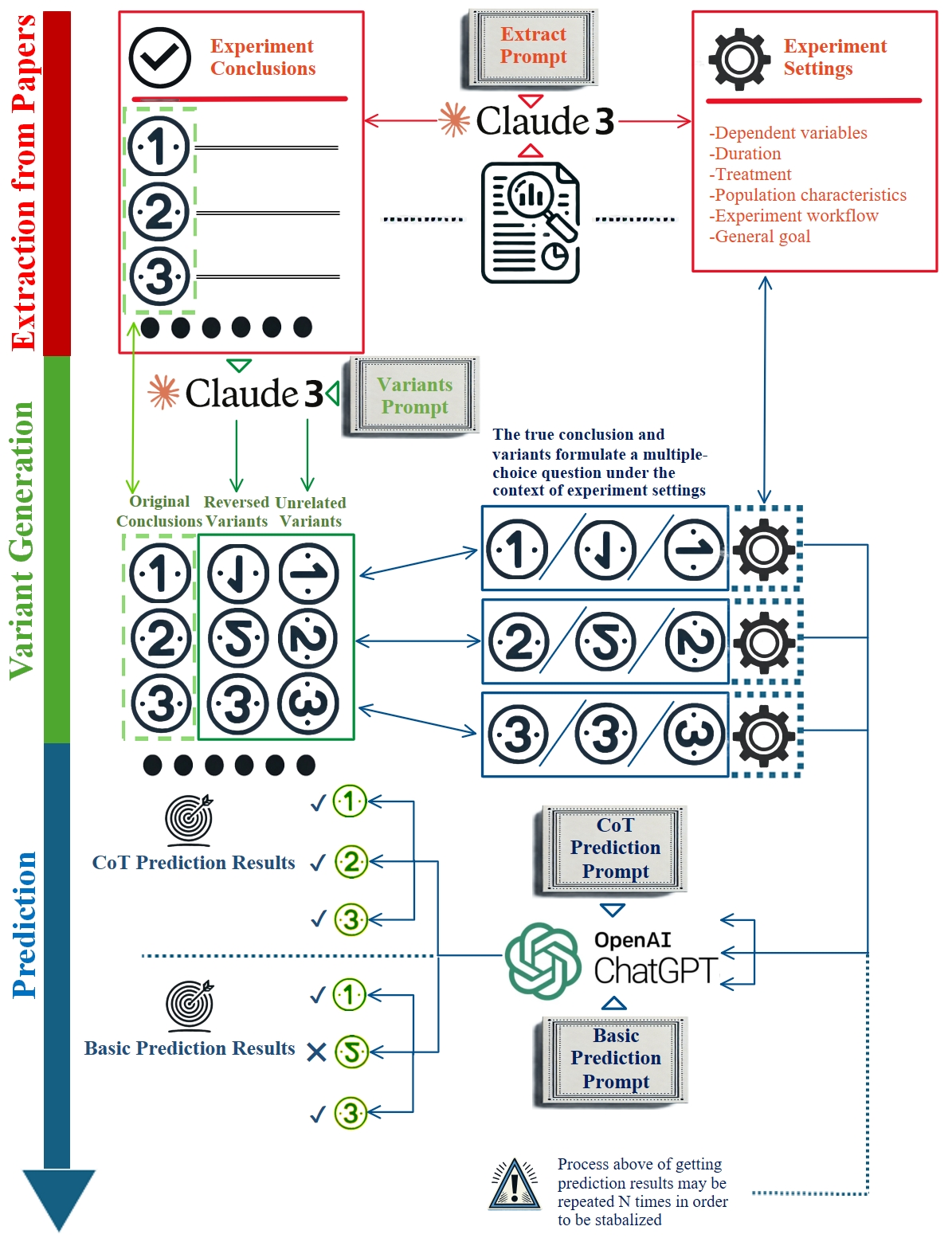}
  \caption{Prediction Framework.}
  \Description{A block diagram showing the three-stage LLM prediction framework: Extraction, Variant Generation, and Prediction, where Claude handles preprocessing and GPT makes final predictions.}

  \label{fig:3}
\end{figure}

We present the details of our automated framework for predicting field experiments in this section. Figure~\ref{fig:3} shows the workflow. Overall, our framework is divided into three stages: Extraction from Papers, Variant Generation, and Prediction. Notably, Claude (Claude-3-opus-20240229) powers all preprocess tasks in the first two stages, whereas GPT completes the prediction at the last stage. We use two different LLMs in different stages to prevent potential data leakage from one another. 

\subsection{Extraction}\label{sec:extraction}

Specifically, the framework uses Claude to extract information related to a field experiment from the selected paper. To realize this, the framework leverages a manually crafted prefixed prompt, which has proven to be efficient for various downstream tasks~\citep{5}. As shown in Figure~\ref{fig:4} (In Appendix~\ref{app:P_extract}), the prompt template contains a placeholder "${Paper}$", an information form consisting of bullet points from "A" to "G", and clear instructions that ask the LLM to extract information according to the form from the paper. As underscored in the prompt, the first six bullet points "A" to "F" are key experimental settings that shape the experiment context, whereas the last point "G" is about conclusions that are true outcomes in the prediction task.

Based on the response from Claude, the framework formulates "Experiment Settings" directly from bullet points "A" to "F", while it polishes point "G" to generate "Experiment Conclusions". Specifically, the raw response regarding point "G" is a paragraph containing multiple conclusions of the field experiment. To separate that paragraph into standalone conclusions, the framework calls a new Claude session, prompting the raw paragraph and related instructions to it, finally getting "Experiment Conclusions" in return. Breaking complex tasks into subtasks improves the performance of LLM-driven workflows~\citep{26}, which is the main reason for completing the extraction and separation of conclusions in different Claude sessions. 


\subsection{Variant Generation}\label{sec:variant generation}

After the experiment settings and conclusions are extracted, the next step is to generate variants based on the true conclusion, since the goal is to see if the LLM could select the true one under distraction. Inspired by Luo et al.~\cite{17}’s prediction of neuroscience results by LLMs, for each conclusion from an experiment, our framework prompts the original conclusion and its two variants to GPT: a reversed variant and an unrelated variant. As a result, the framework will make the prediction by choosing one of three options. 

Specifically, as shown in Figure~\ref{fig:5} (In Appendix~\ref{app:P_vg}), the framework initially prompts the original conclusion to Claude, which follows the instructions to generate the reversed variant of the original conclusion. The reversed variant means that the direction of the conclusion is inverted. For example, if one conclusion is "\textit{receiving housing vouchers reduces quarterly employment rates}," its reversed variant will be "\textit{receiving housing vouchers increases quarterly employment rates}." Next, the framework prompts both the original and reversed conclusions to Claude to generate the unrelated variant, which typically indicates that there is no correlation between entities of interest. Following the same example, the unrelated variant will be "\textit{There is no relationship between receiving housing vouchers and quarterly employment rates}."

\subsection{Prediction}

In the final stage, our framework takes a field experiment's experiment settings, conclusion, and its two variants as input and generates two parallel prediction prompts: basic prediction prompt and Chain-of-Thought (CoT) prompt, which are then prompted to GPT to get predictions by asking GPT to select one conclusion from three conclusions. We also rely on CoT as it has proven to be capable of improving the general performance of LLMs on downstream tasks~\citep{24}. 

Specifically, the basic prediction prompt is shown in Figure~\ref{fig:6} (In Appendix~\ref{app:P_pred}). It consists of a background information section, a question section, and necessary instructions. Specifically, the background information section contains the general goal of the experiment (such as exploring the impact of job training on income), treatments (such as receiving job training or not), experiment duration (such as seven weeks), outcomes (such as income), participant information (such as people seeking jobs in New England), and experiment workflow (such as when and how training was given and outcomes were recorded). All of these were extracted from a target paper, which is the same as bullet points "A" to "F" from Figure~\ref{fig:4} (In Appendix~\ref{app:P_extract}). 

Following the background information, a question section is automatically generated and entered into the templates. Specifically, the original conclusion, its reversed variant, and its unrelated variant are shuffled and substitute the placeholders "option 1", "option 2", and "option 3". Meanwhile, instructions tell GPT that the prediction of conclusions is under the context of the field experimental settings, asking GPT to choose one option from the three options as it deems correct. 

As we are also interested in how CoT would improve such prediction, Figure~\ref{fig:7} (In Appendix~\ref{app:P_pred}) shows the CoT Prediction Prompt, which follows a similar logic as the basic prediction but integrates CoT strategies to boost the performance~\citep{24}. Initially, the framework prompts the experiment settings and three options for a conclusion to GPT, instructing it to think about decisive elements that help choose the correct option. Upon receiving the decisive elements from GPT, the framework prompts GPT to make a selection among three options to get a predicted conclusion. It is worth noting that the entire process is within the same GPT session for a conclusion. 

Although either prompt strategy generates a prediction for a given input, LLMs are stochastic models, meaning that their responses may vary to the same prompt. To handle such randomness in experiment simulations, Leng \& Yuan~\cite{15} set the temperature to 0 and always get fixed responses from synthetic participants of lab experiments, which is a strategy to eliminate the stochasticity of LLMs totally. By contrast, Brand et al.~\cite{4} repeated the same prompt 300 times and used the averaged number as the result of Willingness-to-Pay from customers role-played by LLMs. Here, we take the latter approach by incorporating the stochasticity of LLM outputs since this stochasticity of LLMs is similar to how the same human participant might respond differently when presented with the same instruction~\citep{7}. Specifically, we repeat the same prediction prompt several times and calculate an average accuracy as the final result. For example, if the framework is running the basic prediction, a filled-out prompt based on Figure~\ref{fig:6} (In Appendix~\ref{app:P_pred}) will be repeated a given number of times to get a stable result. The determination of a proper repeat number will be further discussed in the Results section. 

Parameter-wise, no fine-tuning is involved in any stage of the proposed framework, and all parameters of OpenAI API and Anthropic API are set to default. Whereas Horton~\cite{12} harnessed fine-tuned LLMs in lab experimental simulation to better follow instructions, Coda-Forno et al.~\cite{6} simulated several human behaviors by LLMs without fine-tuning. The use of fine-tuned LLMs complicates the reproduction attempts since other researchers don’t have access to the same model in the existing papers~\citep{2}. Additionally, avoiding fine-tuning LLMs saves computing resources and mitigates environmental impacts~\citep{20}, especially when the pretraining of LLMs is enough to make them capable of downstream tasks~\citep{16}.

\subsection{Robustness Checks}\label{sec:Manual Robustness Check}
Given that most steps in the proposed framework are automated, concerns naturally arise regarding the validity of these automated processes. To address these concerns, we conduct three manual screenings to verify the results from extraction (Section~\ref{sec:extraction}) and variant generation (Section~\ref{sec:variant generation}). First, we examine whether the extracted experiment settings (Figure~\ref{fig:4}, Appendix~\ref{app:P_extract}) inadvertently include genuine conclusions. Second, we check whether the extracted conclusions align with those reported in the original papers. Finally, we assess whether the generated variants of the conclusions (Figure~\ref{fig:5}, Appendix~\ref{app:P_vg}) match our expectations.

The first and third screenings revealed no issues. However, the second screening, which examined the alignment of extracted conclusions, identified 377 conclusions as either incomprehensible or nonexistent in the original texts. Consequently, after the three screenings, 1261 conclusions and 276 papers remained and were deemed valid for our purposes.

Additionally, 86 out of 1261 conclusions were dequantified, as existing studies suggest that predicting the magnitude of experimental outcomes remains challenging at this stage~\citep{18}. Given our focus on predicting the direction of experimental conclusions rather than the magnitude, conclusions specifying precise numerical treatment effects were reformulated in a dequantified manner. For instance, a conclusion such as "Job training increases income by 30\%" was revised to "Job training increases income."

\section{Results}
In this section, we test our framework on 276 field experiments that contain a total of 1261 conclusions and discuss the performance.

We use Conclusion Accuracy and Paper Accuracy to evaluate the prediction performance under different settings.  As shown in Figure~\ref{fig:3}, the framework generates a prompt for each conclusion. The generated prompt either follows the template in Figure~\ref{fig:6} or~\ref{fig:7} (In Appendix~\ref{app:P_pred}), depending on which prompt strategy (basic or CoT) the framework applies. Each prompt instructs GPT to output a predicted conclusion. If the predicted conclusion matches the true conclusion, that attempt is counted as correct. As aforementioned, the framework repeats such an attempt for a set number of times for each conclusion to get a stable result. Therefore, we define \textit{Conclusion Accuracy} as the percentage of correct predictions among a set number of attempts (Equation~\ref{eq:1}). Given that a field experiment may contain multiple conclusions, \textit{Paper Accuracy} is the average of all \textit{Conclusion Accuracy} within a paper (Equation~\ref{eq:2}). 
\begin{equation}
\label{eq:1}
\text{Conclusion Accuracy} = \frac{\text{Num of Correct Predictions}}{\text{Num of Predictions}} \times 100\%
\end{equation}

\begin{equation}
\label{eq:2}
\text{Paper Accuracy} = \frac{1}{N}\sum_{i=1}^{N}\text{Conclusion Accuracy}_i
\end{equation}

\subsection{Prediction Performance Overview}
\begin{table}[h]
  \caption{Prediction Accuracy under Different Repeats}
  \label{tab:1}
  \footnotesize
  \centering
  \begin{tabular}{lcccc}
    \toprule
    \multirow{2}{*}{GPT-4 Turbo} & \multicolumn{2}{c}{Basic} & \multicolumn{2}{c}{CoT} \\
    \cmidrule(lr){2-3} \cmidrule(lr){4-5}
     & Conclusion Acc. & Paper Acc. & Conclusion Acc. & Paper Acc. \\
    \midrule
    Repeat = 10 & 66\% & 66\% & 76\% & 76\% \\
    Repeat = 20 & 66\% & 66\% & 76\% & 76\% \\
    Repeat = 30 & 65\% & 66\% & 76\% & 76\% \\
    \bottomrule
  \end{tabular}
  \vspace{0.5ex}
  \raggedright \footnotesize Sample size: 1261 conclusions from 276 papers. In the experiment we use gpt-3.5-turbo-0125, gpt-4-turbo-2024-04-09, and gpt-4o-2024-11-20. 
\end{table}

\begin{table}[h]
  \caption{Prediction Accuracy under 20 Repeats by GPT Models}
  \label{tab:2}
  \footnotesize
  \centering
  \begin{tabular}{lcccc}
    \toprule
    \multirow{2}{*}{Model} & \multicolumn{2}{c}{Basic} & \multicolumn{2}{c}{CoT} \\
    \cmidrule(lr){2-3} \cmidrule(lr){4-5}
     & Conclusion Acc. & Paper Acc. & Conclusion Acc. & Paper Acc. \\
    \midrule
    GPT-3.5 Turbo & 61\% & 61\% & 68\% & 67\% \\
    GPT-4 Turbo   & 66\% & 66\% & 76\% & 76\% \\
    GPT-4o        & 75\% & 75\% & 78\% & 78\% \\
    \bottomrule
  \end{tabular}
  \vspace{0.5ex}
  \raggedright \footnotesize Sample size: 1261 conclusions from 276 papers. In the experiment we use gpt-3.5-turbo-0125, gpt-4-turbo-2024-04-09, and gpt-4o-2024-11-20. 
\end{table}

Table~\ref{tab:1} reports two types of accuracy for both strategies under different repeats by GPT4-turbo. The best results for the basic strategy are obtained under 10 repeats, which are 66\% for both conclusion accuracy and for paper accuracy. By comparison, the CoT results are the same 76\% for both strategies and different repeats, which are generally 10\% points higher than basic results and aligns with prior literature on boosting LLM performance by CoT~\citep{24}. Meanwhile, another key observation is that the results show a certain degree of invariability to repeat numbers. Specifically, the results across different numbers of attempts don’t seem to change under either the basic or CoT strategy. Given that more repeats result in higher time and monetary cost, we chose 20 repeats for further evaluation.

Table~\ref{tab:2} reports results by different GPT models with 20 attempts. One key observation from Table~\ref{tab:2} is that prediction performance steadily improves as LLMs iterate, though the rate of improvement decreases over time. According to Table~\ref{tab:2}, the best result under 20 repeats is achieved by GPT4o under CoT prompt strategy, which is a conclusion accuracy of 78\% and a paper accuracy of 78\%. Specifically, on average, GPT4o is able to predict a conclusion in 78\% of the 20 repeated prediction attempts, and it also predicts 78\% of the outcomes correctly for each paper. While the conclusion accuracy of GPT4o under CoT strategy is 10 percentage points higher than the CoT result of GPT3-turbo, it is only two percentage points higher than the CoT result of GPT4-turbo, indicating the improvement from model iteration becomes harder for this task. 

Another interesting finding from Table~\ref{tab:2} is the boosting effect of CoT on performance varies on models for the experiment conclusion prediction task. Specifically, the largest improvement is 10 percentage points on GPT4-turbo, while the least improvement is three percentage points on GPT4o. However, it’s unsafe to conclude that CoT boosting is weaker on newer models since CoT improves the accuracy by seven percentage points on the oldest model tested, GPT3-turbo.

In summary, our findings indicate that incorporating a CoT strategy generally enhances predictive performance, whereas increasing the number of repetitions does not produce a significant effect. Furthermore, iterative refinements to LLMs consistently improve performance, although the rate of improvement diminishes over time. Finally, the performance gains attributable to CoT appear to be model-sensitive.

\subsection{Data Memorization}

\begin{figure}[h]
  \centering
  \includegraphics[width=\linewidth]{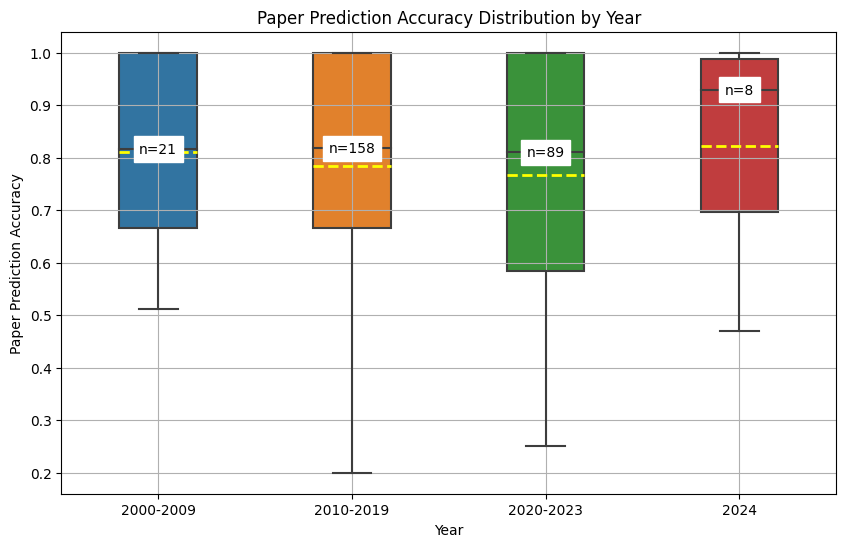}
  \caption{Paper Accuracy by Year. }
  \Description{A bar chart showing paper-level prediction accuracy grouped by publication year, used to assess potential data memorization by comparing older and newer field experiments.}

  \label{fig:11}
\end{figure}

Data memorization is a common concern in simulating experiments with LLMs. If the results given by the LLM are from its memory of training data instead of reasoning, the proposed idea has no instructional value as pilot testing for field experiments~\citep{12}. As revealed in its documents, the training data cutoff of gpt-4o-2024-11-20 is October 2023. Therefore, it is reasonable to assume that the model would not have seen the field experiments papers appeared in 2024. In other words, field experiments published after 2024 would be less likely to be included in the training data. 

To text our framework's robustness in light of data memorization, we split the papers by year. Figure~\ref{fig:11} shows that the paper accuracy is actually higher on papers 2024 than any other years. This is notable because if the prediction results are driven by LLMs' memorization of its training data, the framework's performance in unseen papers in 2024 would be poorer, indicating data memorization of less concern to our framework. 






\begin{figure}[t]
  \centering
  \begin{subfigure}[t]{\linewidth}
    \centering
    \includegraphics[width=0.95\linewidth]{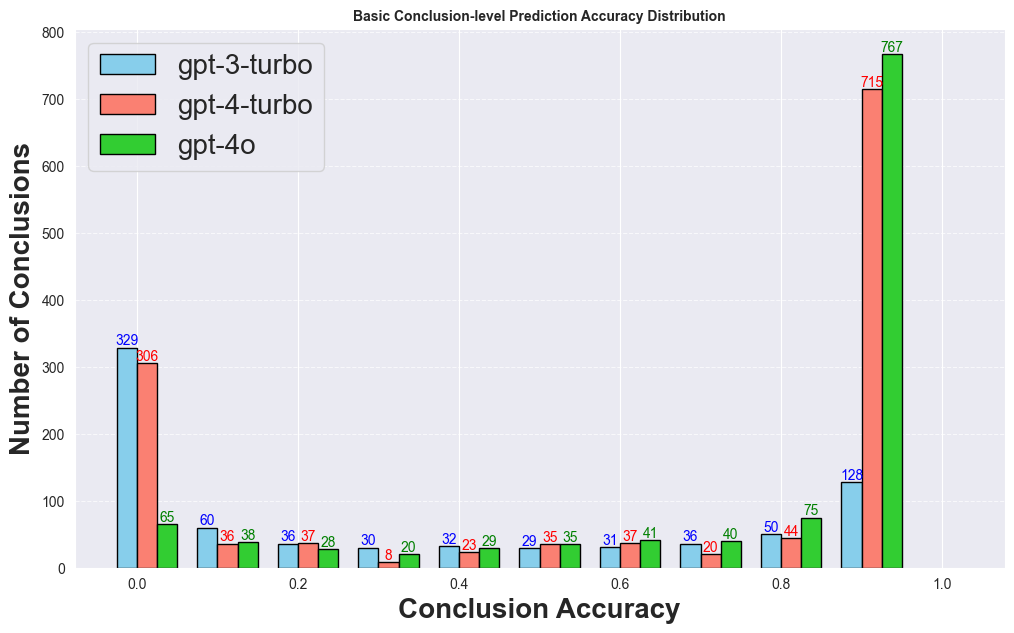}
    \caption{Basic Conclusion Prediction Accuracy}
    \label{fig:8a}
  \end{subfigure}
  \vspace{1ex}
  \begin{subfigure}[t]{\linewidth}
    \centering
    \includegraphics[width=0.95\linewidth]{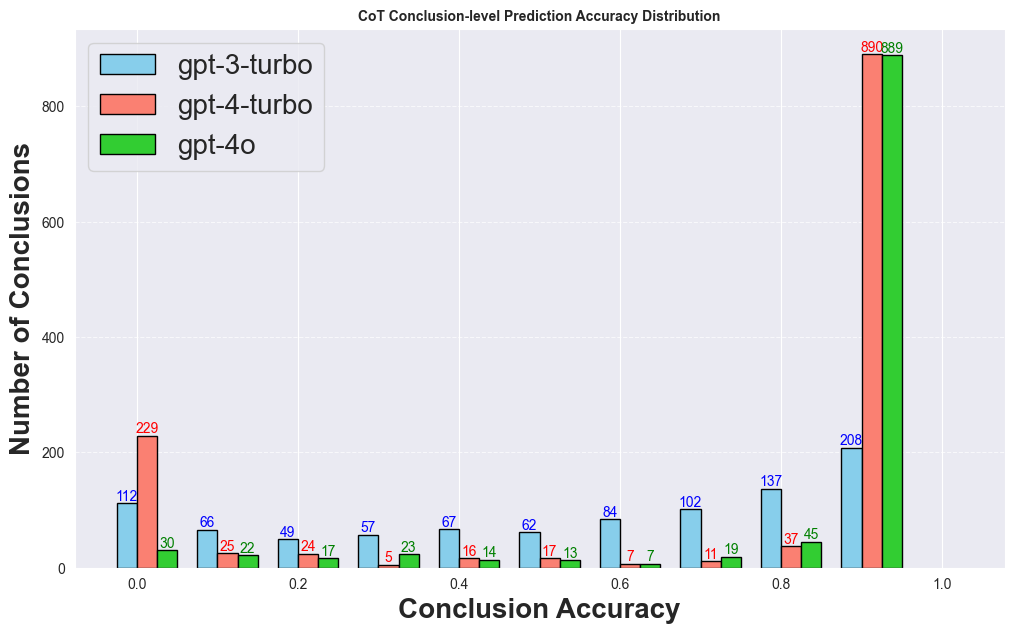}
    \caption{CoT Conclusion Prediction Accuracy}
    \label{fig:8b}
  \end{subfigure}
  \caption{Conclusion Prediction Accuracy Distribution}
  \Description{Two vertically stacked bar charts showing conclusion-level prediction accuracy for basic prompting and chain-of-thought prompting.}
  \label{fig:8}
\end{figure}

\begin{figure}[t]
  \centering
  \begin{subfigure}[t]{\linewidth}
    \centering
    \includegraphics[width=0.95\linewidth]{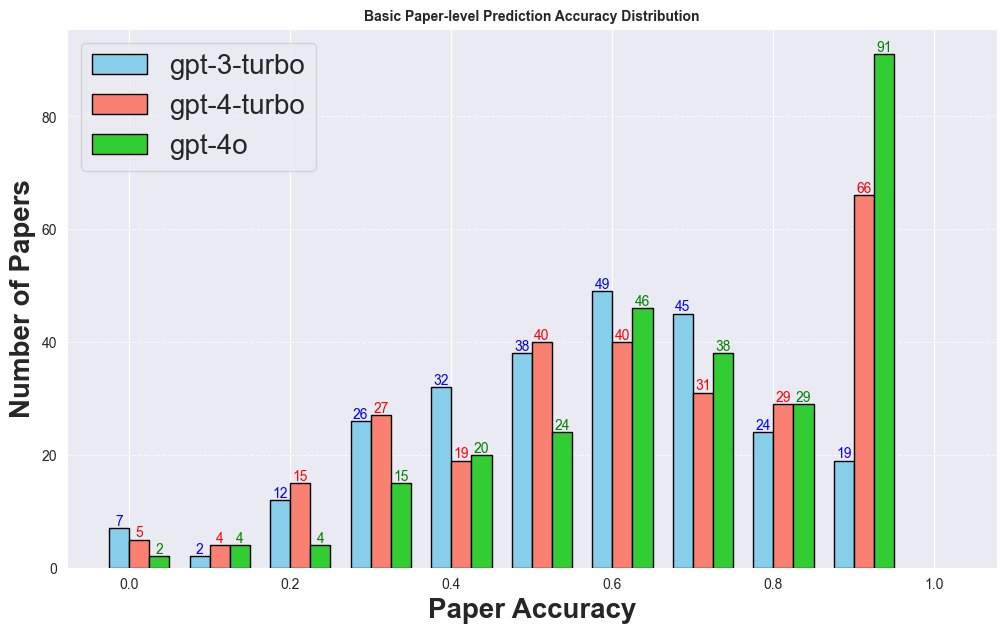}
    \caption{Basic Paper Prediction Accuracy}
    \label{fig:9a}
  \end{subfigure}
  \vspace{1ex}
  \begin{subfigure}[t]{\linewidth}
    \centering
    \includegraphics[width=0.95\linewidth]{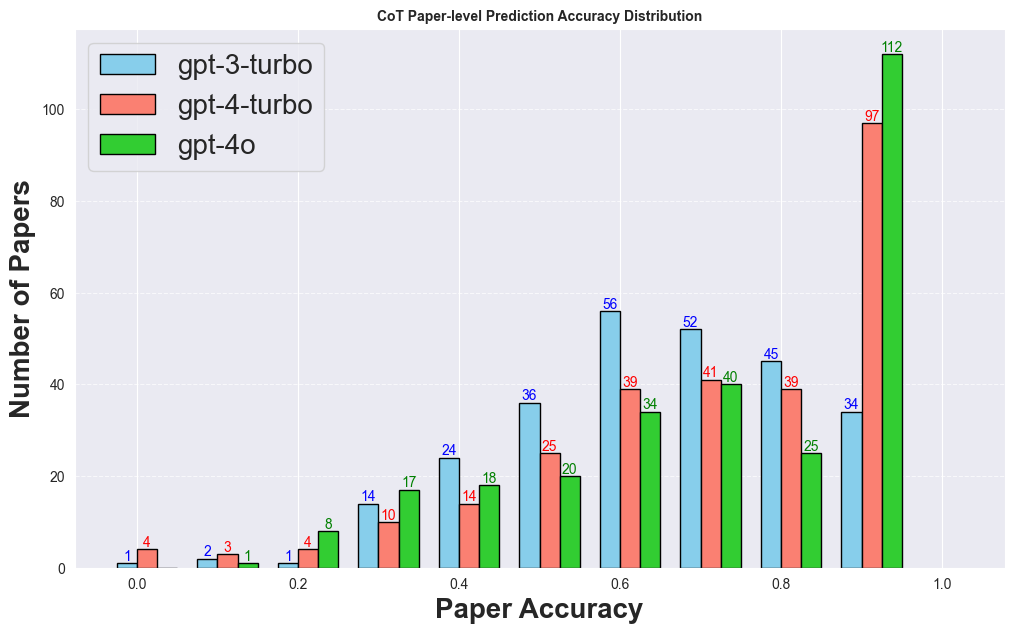}
    \caption{CoT Paper Prediction Accuracy}
    \label{fig:9b}
  \end{subfigure}
  \caption{Paper Prediction Accuracy Distribution}
  \Description{Two vertically stacked bar charts showing prediction-level prediction accuracy for basic prompting and chain-of-thought prompting.}
  \label{fig:9}
\end{figure}

\subsection{Examination of Distributions of Results}\label{sec:Anomalous Distributions of Results}

To further examine the prediction performance of our proposed framework, we plot the distributions of accuracy results reported in Table~\ref{tab:2}, based on different GPT models and two prompt strategies under 20 repeats (Figure~\ref{fig:8} and~\ref{fig:9}). 

As shown in Figure~\ref{fig:8a}, the conclusion accuracy of all three models (gpt-3, gpt4-turbo and gpt-4o) exhibits bimodal patterns, with peaks concentrated toward both the lower and upper extremes (a U-shaped distribution). Specifically, there are significant concentrations of samples in the 90\%-100\% accuracy range and another notable cluster in the 0\%-10\% accuracy range. This indicates that the model's performance is highly polarized, where certain conclusions are predicted with near-perfect accuracy, while others are almost wrongly predicted entirely. Interestingly, by applying CoT strategy (Figure~\ref{fig:8b}), the concentration in the 0\%-10\% accuracy range is mitigated, while the concentration in the 90\%-100\% accuracy range deepens, resulting from the performance boosting by CoT. Model-wise, the U-shaped distributions for the earlier model (gpt-3-turbo) are milder compared to recent models (gpt-4-turbo and gpt-4o), probably resulting from the performance limitation of earlier models. 

Similar patterns also exist in Figure~\ref{fig:9} to show the paper's accuracy. Distributions for all models are skewed. Such a skewness is more pronounced in gpt-4-turbo and gpt-4o, with samples concentrating in the upper extreme. This indicates that the framework correctly predicts all conclusions from certain tested experiments. Additionally,  CoT generally increases the degree of  skewness(Figure~\ref{fig:9b}), aligning with its boosting effect on LLMs' performance.

Inspired by the bimodal and skewed results (Section~\ref{sec:Anomalous Distributions of Results}), we closely examined the topics of each conclusion based on its experimental context. Leveraging LLMs' ability in annotating~\citep{temp2},  we prompted Claude to label topic components of each conclusion under the context of the experiment, as shown in Figure~\ref{fig:10a}. To ensure selected topics could grasp the reason behind the abnomalities found previously (Figure~\ref{fig:8} and Figure~\ref{fig:9}), we included topics where LLMs' responses are biased~\citep{bias1,bias2,bias3,bias4,bias5}, including gender, ethnicity, social norms, ethical dilemmas, age, socioeconomic status, and other topics. As a result, each conclusion was represented by a vector of percentages summing to 100\%, where each percentage indicated the degree to which the context was associated with a particular topic. Furthermore, as the sentiment bias also affects the generated content of LLMs on top of topics bias~\citep{sentiment1}, we used Claude as the sentiment analysis tool to label each conclusion, deciding the sentiment of each conclusion, either positive, negative, or neutral (Figure~\ref{fig:10b}, Appendix~\ref{app:P_label}). In addition to regular sentiments, for gender-related conclusions, Claude also labeled each of them with a gender favorability tag, as LLMs may favor and tend to generate pro-female content~\citep{gender1}. Specifically, if the context of a conclusion relates to gender, Claude would further judge if the context is favorable to females or detrimental to males, the opposite, or neutral, as shown in Figure~\ref{fig:10c}. Finally, all variables acquired from the labeling process are summarized in Table~\ref{tab:variables} (In Appendix~\ref{app:P_rv}).

To study the impact of topic components and sentiments on LLMs' performance in experimental prediction, we constructed a regression model (Equation~\ref{eq:regression_model}). Table~\ref{tab:regression_results} reports six regression results by three GPT models and two prompt strategies under 20 repeats, which corresponds with the conclusion accuracy results in Table~\ref{tab:2}. Notably, the sample size, which is the number of conclusions, decreases from 1261 to 955 because the content filtering policy prevents Claude from generating labels in the labeling process (Figure~\ref{fig:10a},~\ref{fig:10b},~\ref{fig:10c}, Appendix~\ref{app:P_label}). 

\begin{strip}
\centering
\begin{minipage}{0.95\textwidth}
\begin{equation}
\label{eq:regression_model}
\begin{aligned}
\text{Conclusion Accuracy} =\; & \beta_0 + \beta_1 (\text{Gender}) + \beta_2 (\text{Ethnicity}) + \beta_3 (\text{Social norms}) \\
& + \beta_4 (\text{Ethical dilemmas}) + \beta_5 (\text{Age}) + \beta_6 (\text{Socioeconomic status}) \\
& + \beta_7 (\text{Other topics}) + \beta_8 (\text{Gender} \times \text{Favorability}) \\
& + \beta_9 (\text{Ethnicity} \times \text{Sentiment}) + \beta_{10} (\text{Social norms} \times \text{Sentiment}) \\
& + \beta_{11} (\text{Ethical dilemmas} \times \text{Sentiment}) + \beta_{12} (\text{Age} \times \text{Sentiment}) \\
& + \beta_{13} (\text{Socioeconomic status} \times \text{Sentiment}) + \beta_{14} (\text{Other topics} \times \text{Sentiment}) \\
& + \epsilon
\end{aligned}
\end{equation}
\end{minipage}
\captionof{figure}{Full specification of the regression model used to examine the effect of topic–sentiment interactions on prediction accuracy.}
\label{fig:regression-model}
\end{strip}

There are several key findings from regression results (Table~\ref{tab:regression_results}). First, the interaction effects between certain topics and sentiments significantly affect the performance, though there is no significant evidence that topic components alone may affect LLMs' ability in predicting experimental conclusions. Specifically, sentiment interactions with complex social issues such as ethnicity, social norms, age, and other topics are significantly positive, while sentiment interactions with ethical dilemmas is significantly negative. For example, the regression of gpt4o and basic prompt (Table~\ref{tab:regression_results} (1)) shows that sentiment interactions with ethnicity and other topics are significantly positive, suggesting if the context of a conclusions is positive and relates to ethnicity, LLMs tend to predict it correctly, while sentiment interaction with ethical dilemmas is significantly negative, suggesting that if the context of a conclusion is negative and relates to ethical dilemmas, LLMs tend to avoid it, resulting in a wrong prediction. Surprisingly, gender, which is a topic component widely perceived to be biased in LLMs~\citep{bias5, gender1}, seems not to be a biased topic in LLM-based experiment prediction. Favorability interaction with gender is only significantly positive in gpt4o prediction under CoT, suggesting that gpt4o inclines to predict pro-female conclusions and avoid pro-male conclusions under CoT strategy only in one out of six scenarios. This is also the case for the age topic, as the coefficient of its interaction is positive only in one regression. Another interesting finding is that estimated coefficients of socioeconomic status and its interaction terms are never significant according to the regression results, suggesting that LLM prediction is unbiased to conclusions related to this topic. 

Second, though the directions of estimated coefficients are stable, the significant levels of the same interaction term vary across regressions due to the refinement from model iteration and prompt strategies. Specifically, the estimated coefficients of the ethnicity's interaction term are significant across all gpt models, suggesting that the iteration LLMs didn't fix this bias. By comparison, the coefficients of the interaction of social norms are no longer significant since gpt4-turbo, suggesting that this bias might have been fixed. 

Third, applying CoT strategy might mitigate the impact of sentiment interaction with certain topics. According to the regression results in Table~\ref{tab:regression_results} (1) and Table~\ref{tab:regression_results} (2), implementing CoT turns some significant estimated coefficients of interaction terms into nonsignificant. However, this is not the case for gpt3-turbo model, as CoT brings more significant coefficients.





\section{Conclusion}
In this paper, we propose an LLM-powered framework that automatically extracts information from existing papers and predicts field experimental conclusions. To the best of our knowledge, this is the first paper to provide an automated framework for predicting such conclusions for field experiments. Rather than merely introducing the framework, our work also examines its fidelity on a large scale of samples and achieves a considerable accuracy of 78\%. Furthermore, the paper discovers that incorporating a CoT strategy generally enhances predictive performance in this scenario, whereas the performance gains attributable to CoT appear to be model-dependent. Furthermore, iterative refinements to LLMs consistently improve performance, although the rate of improvement diminishes over time. 

Interestingly, the paper also finds that the distributions of prediction accuracy are either bimodal or negatively skewed, with a large number of samples concentrating on the two extremes. To explore this phenomenon, the paper regresses conclusion prediction accuracy on topic components and sentiments, revealing that interaction effects between certain topics and sentiments could affect the LLMs' prediction performance in this task. 

Taken together, these findings underscore the potential of the LLM-driven frameworks in advancing automated predictions for field experiments while also clarifying the practical constraints that guide their effective use.

\begin{strip}
  \centering
  \captionof{table}{OLS Regression Results for Conclusion Accuracy}
  \label{tab:regression_results}
  \scriptsize
  \centering
  \resizebox{\textwidth}{!}{ 
  \begin{tabular}{lcccccc}
    \toprule
    & (1) & (2) & (3) & (4) & (5) & (6) \\
    Variable & GPT-4o Basic & GPT-4o CoT & GPT-4 Turbo Basic & GPT-4 Turbo CoT & GPT-3.5 Turbo Basic & GPT-3.5 Turbo CoT \\
    \midrule
    Constant & 0.919 (0.721) & -0.240 (0.929) & 3.397 (0.249) & 1.860 (0.498) & 4.026 (0.182) & 2.820 (0.218) \\
    Gender & -0.138 (0.957) & 1.006 (0.710) & -2.932 (0.320) & -1.084 (0.693) & -3.628 (0.230) & -2.245 (0.327) \\
    Ethnicity & -0.342 (0.893) & 1.103 (0.679) & -2.903 (0.318) & -1.347 (0.620) & -3.485 (0.242) & -2.209 (0.328) \\
    Social Norms & -0.124 (0.962) & 1.129 (0.677) & -2.483 (0.402) & -1.241 (0.653) & -3.607 (0.235) & -2.251 (0.328) \\
    Ethical Dilemmas & -0.729 (0.779) & 0.908 (0.738) & -2.821 (0.341) & -0.600 (0.828) & -3.156 (0.298) & -2.068 (0.369) \\
    Age & -0.151 (0.954) & 0.838 (0.757) & -2.868 (0.332) & -1.233 (0.655) & -3.366 (0.267) & -2.231 (0.332) \\
    Socioeconomic Status & -0.006 (0.998) & 1.071 (0.692) & -2.747 (0.352) & -1.108 (0.687) & -3.486 (0.249) & -2.120 (0.355) \\
    Other Topics & -0.218 (0.933) & 0.955 (0.724) & -2.880 (0.329) & -1.263 (0.646) & -3.423 (0.257) & -2.225 (0.331) \\
    Gender × Favorability & -0.031 (0.897) & 0.478* (0.057) & 0.125 (0.647) & 0.089 (0.727) & -0.313 (0.264) & 0.202 (0.341) \\
    Ethnicity × Sentiment & 0.349** (0.027) & 0.242 (0.142) & 0.519*** (0.004) & 0.439*** (0.009) & 0.126 (0.493) & 0.323** (0.021) \\
    Social Norms × Sentiment & 0.112 (0.615) & 0.121 (0.602) & 0.191 (0.453) & 0.355 (0.134) & 0.692*** (0.008) & 0.333* (0.092) \\
    Ethical Dilemmas × Sentiment & -0.605** (0.028) & -0.318 (0.269) & -0.544* (0.083) & -0.810*** (0.006) & -0.525 (0.103) & -0.346 (0.156) \\
    Age × Sentiment & 0.225 (0.247) & -0.187 (0.357) & 0.447** (0.044) & 0.176 (0.394) & 0.114 (0.617) & 0.122 (0.479) \\
    Socioeconomic Status × Sentiment & -0.011 (0.893) & -0.082 (0.345) & 0.101 (0.289) & 0.124 (0.161) & 0.051 (0.599) & 0.062 (0.401) \\
    Other Topics × Sentiment & 0.126* (0.053) & 0.115* (0.091) & 0.081 (0.275) & 0.199*** (0.004) & -0.019 (0.800) & 0.126** (0.029) \\
    \midrule
    $R^2$ & 0.041 & 0.017 & 0.065 & 0.076 & 0.033 & 0.079 \\
    Adj. $R^2$ & 0.027 & 0.002 & 0.051 & 0.062 & 0.019 & 0.065 \\
    F-statistic & 2.859 & 1.138 & 4.660 & 5.542 & 2.323 & 5.769 \\
    Prob. (F-statistic) & 0.0003 & 0.319 & 0.000 & 0.000 & 0.004 & 0.000 \\
    Observations & 955 & 955 & 955 & 955 & 955 & 955 \\
    \bottomrule
  \end{tabular}
  } 
  \vspace{1ex}
  \raggedright \scriptsize
  Note: Coefficients with p-values in parentheses. *** $p<0.01$, ** $p<0.05$, * $p<0.1$. Also, for interaction terms, since our definition of pro-female content includes negative sentiment when the context is detrimental to the male, it's not sufficient to represent the gender favorability label with regular sentiment label. Therefore, for the gender topic, we explored its interaction with gender favorability label, while for the rest of the topics, we explored their interactions with sentiment.
\end{strip}

\bibliographystyle{ACM-Reference-Format}
\bibliography{sample-base}


\begin{thebibliography}{50}


\ifx \showCODEN    \undefined \def \showCODEN     #1{\unskip}     \fi
\ifx \showISBNx    \undefined \def \showISBNx     #1{\unskip}     \fi
\ifx \showISBNxiii \undefined \def \showISBNxiii  #1{\unskip}     \fi
\ifx \showISSN     \undefined \def \showISSN      #1{\unskip}     \fi
\ifx \showLCCN     \undefined \def \showLCCN      #1{\unskip}     \fi
\ifx \shownote     \undefined \def \shownote      #1{#1}          \fi
\ifx \showarticletitle \undefined \def \showarticletitle #1{#1}   \fi
\ifx \showURL      \undefined \def \showURL       {\relax}        \fi
\providecommand\bibfield[2]{#2}
\providecommand\bibinfo[2]{#2}
\providecommand\natexlab[1]{#1}
\providecommand\showeprint[2][]{arXiv:#2}

\bibitem[Abid et~al\mbox{.}(2021)]%
        {bias2}
\bibfield{author}{\bibinfo{person}{Abubakar Abid}, \bibinfo{person}{Maheen Farooqi}, {and} \bibinfo{person}{James Zou}.} \bibinfo{year}{2021}\natexlab{}.
\newblock \showarticletitle{Persistent anti-muslim bias in large language models}. In \bibinfo{booktitle}{\emph{Proceedings of the 2021 AAAI/ACM Conference on AI, Ethics, and Society}}. \bibinfo{pages}{298--306}.
\newblock


\bibitem[Aher et~al\mbox{.}(2023)]%
        {1}
\bibfield{author}{\bibinfo{person}{Gati~V Aher}, \bibinfo{person}{Rosa~I Arriaga}, {and} \bibinfo{person}{Adam~Tauman Kalai}.} \bibinfo{year}{2023}\natexlab{}.
\newblock \showarticletitle{Using large language models to simulate multiple humans and replicate human subject studies}. In \bibinfo{booktitle}{\emph{International Conference on Machine Learning}}. PMLR, \bibinfo{pages}{337--371}.
\newblock


\bibitem[Anastasi et~al\mbox{.}(2024)]%
        {tempo2}
\bibfield{author}{\bibinfo{person}{Joyce~K Anastasi}, \bibinfo{person}{Bernadette Capili}, \bibinfo{person}{Margaret Norton}, \bibinfo{person}{Donald~J McMahon}, {and} \bibinfo{person}{Karen Marder}.} \bibinfo{year}{2024}\natexlab{}.
\newblock \showarticletitle{Recruitment and retention of clinical trial participants: understanding motivations of patients with chronic pain and other populations}.
\newblock \bibinfo{journal}{\emph{Frontiers in Pain Research}}  \bibinfo{volume}{4} (\bibinfo{year}{2024}), \bibinfo{pages}{1330937}.
\newblock


\bibitem[Bail(2024)]%
        {2}
\bibfield{author}{\bibinfo{person}{Christopher~A Bail}.} \bibinfo{year}{2024}\natexlab{}.
\newblock \showarticletitle{Can Generative AI improve social science?}
\newblock \bibinfo{journal}{\emph{Proceedings of the National Academy of Sciences}} \bibinfo{volume}{121}, \bibinfo{number}{21} (\bibinfo{year}{2024}), \bibinfo{pages}{e2314021121}.
\newblock


\bibitem[Berge et~al\mbox{.}(2015)]%
        {if5}
\bibfield{author}{\bibinfo{person}{Lars Ivar~Oppedal Berge}, \bibinfo{person}{Kjetil Bjorvatn}, {and} \bibinfo{person}{Bertil Tungodden}.} \bibinfo{year}{2015}\natexlab{}.
\newblock \showarticletitle{Human and financial capital for microenterprise development: Evidence from a field and lab experiment}.
\newblock \bibinfo{journal}{\emph{Management Science}} \bibinfo{volume}{61}, \bibinfo{number}{4} (\bibinfo{year}{2015}), \bibinfo{pages}{707--722}.
\newblock


\bibitem[Brand et~al\mbox{.}(2023)]%
        {4}
\bibfield{author}{\bibinfo{person}{James Brand}, \bibinfo{person}{Ayelet Israeli}, {and} \bibinfo{person}{Donald Ngwe}.} \bibinfo{year}{2023}\natexlab{}.
\newblock \showarticletitle{Using GPT for market research}.
\newblock \bibinfo{journal}{\emph{Harvard Business School Marketing Unit Working Paper}} \bibinfo{number}{23-062} (\bibinfo{year}{2023}).
\newblock


\bibitem[Brown et~al\mbox{.}(2020)]%
        {5}
\bibfield{author}{\bibinfo{person}{Tom Brown}, \bibinfo{person}{Benjamin Mann}, \bibinfo{person}{Nick Ryder}, \bibinfo{person}{Melanie Subbiah}, \bibinfo{person}{Jared~D Kaplan}, \bibinfo{person}{Prafulla Dhariwal}, \bibinfo{person}{Arvind Neelakantan}, \bibinfo{person}{Pranav Shyam}, \bibinfo{person}{Girish Sastry}, \bibinfo{person}{Amanda Askell}, {et~al\mbox{.}}} \bibinfo{year}{2020}\natexlab{}.
\newblock \showarticletitle{Language models are few-shot learners}.
\newblock \bibinfo{journal}{\emph{Advances in neural information processing systems}}  \bibinfo{volume}{33} (\bibinfo{year}{2020}), \bibinfo{pages}{1877--1901}.
\newblock


\bibitem[Chen et~al\mbox{.}(2025)]%
        {if10}
\bibfield{author}{\bibinfo{person}{Yang Chen}, \bibinfo{person}{Samuel~N Kirshner}, \bibinfo{person}{Anton Ovchinnikov}, \bibinfo{person}{Meena Andiappan}, {and} \bibinfo{person}{Tracy Jenkin}.} \bibinfo{year}{2025}\natexlab{}.
\newblock \showarticletitle{A manager and an AI walk into a bar: does ChatGPT make biased decisions like we do?}
\newblock \bibinfo{journal}{\emph{Manufacturing \& Service Operations Management}} (\bibinfo{year}{2025}).
\newblock


\bibitem[Chu and Shen(2006)]%
        {if9}
\bibfield{author}{\bibinfo{person}{Leon~Yang Chu} {and} \bibinfo{person}{Zuo-Jun~Max Shen}.} \bibinfo{year}{2006}\natexlab{}.
\newblock \showarticletitle{Agent competition double-auction mechanism}.
\newblock \bibinfo{journal}{\emph{Management Science}} \bibinfo{volume}{52}, \bibinfo{number}{8} (\bibinfo{year}{2006}), \bibinfo{pages}{1215--1222}.
\newblock


\bibitem[Coda-Forno et~al\mbox{.}(2024)]%
        {6}
\bibfield{author}{\bibinfo{person}{Julian Coda-Forno}, \bibinfo{person}{Marcel Binz}, \bibinfo{person}{Jane~X Wang}, {and} \bibinfo{person}{Eric Schulz}.} \bibinfo{year}{2024}\natexlab{}.
\newblock \showarticletitle{CogBench: a large language model walks into a psychology lab}.
\newblock \bibinfo{journal}{\emph{arXiv preprint arXiv:2402.18225}} (\bibinfo{year}{2024}).
\newblock


\bibitem[Cui et~al\mbox{.}(2024)]%
        {auto1}
\bibfield{author}{\bibinfo{person}{Ziyan Cui}, \bibinfo{person}{Ning Li}, {and} \bibinfo{person}{Huaikang Zhou}.} \bibinfo{year}{2024}\natexlab{}.
\newblock \showarticletitle{Can AI Replace Human Subjects? A Large-Scale Replication of Psychological Experiments with LLMs}.
\newblock \bibinfo{journal}{\emph{arXiv preprint arXiv:2409.00128}} (\bibinfo{year}{2024}).
\newblock


\bibitem[Demszky et~al\mbox{.}(2023)]%
        {7}
\bibfield{author}{\bibinfo{person}{Dorottya Demszky}, \bibinfo{person}{Diyi Yang}, \bibinfo{person}{David~S Yeager}, \bibinfo{person}{Christopher~J Bryan}, \bibinfo{person}{Margarett Clapper}, \bibinfo{person}{Susannah Chandhok}, \bibinfo{person}{Johannes~C Eichstaedt}, \bibinfo{person}{Cameron Hecht}, \bibinfo{person}{Jeremy Jamieson}, \bibinfo{person}{Meghann Johnson}, {et~al\mbox{.}}} \bibinfo{year}{2023}\natexlab{}.
\newblock \showarticletitle{Using large language models in psychology}.
\newblock \bibinfo{journal}{\emph{Nature Reviews Psychology}} \bibinfo{volume}{2}, \bibinfo{number}{11} (\bibinfo{year}{2023}), \bibinfo{pages}{688--701}.
\newblock


\bibitem[Dillion et~al\mbox{.}(2023)]%
        {8}
\bibfield{author}{\bibinfo{person}{Danica Dillion}, \bibinfo{person}{Niket Tandon}, \bibinfo{person}{Yuling Gu}, {and} \bibinfo{person}{Kurt Gray}.} \bibinfo{year}{2023}\natexlab{}.
\newblock \showarticletitle{Can AI language models replace human participants?}
\newblock \bibinfo{journal}{\emph{Trends in Cognitive Sciences}} \bibinfo{volume}{27}, \bibinfo{number}{7} (\bibinfo{year}{2023}), \bibinfo{pages}{597--600}.
\newblock


\bibitem[Du et~al\mbox{.}(2020)]%
        {if6}
\bibfield{author}{\bibinfo{person}{Ninghua Du}, \bibinfo{person}{Lingfang Li}, \bibinfo{person}{Tian Lu}, {and} \bibinfo{person}{Xianghua Lu}.} \bibinfo{year}{2020}\natexlab{}.
\newblock \showarticletitle{Prosocial compliance in P2P lending: A natural field experiment}.
\newblock \bibinfo{journal}{\emph{Management Science}} \bibinfo{volume}{66}, \bibinfo{number}{1} (\bibinfo{year}{2020}), \bibinfo{pages}{315--333}.
\newblock


\bibitem[Du and Zhang(2024)]%
        {ijk5}
\bibfield{author}{\bibinfo{person}{Xiaocong Du} {and} \bibinfo{person}{Haipeng Zhang}.} \bibinfo{year}{2024}\natexlab{}.
\newblock \showarticletitle{For the Misgendered Chinese in Gender Bias Research: Multi-Task Learning with Knowledge Distillation for Pinyin Name-Gender Prediction}.
\newblock \bibinfo{journal}{\emph{arXiv preprint arXiv:2405.06221}} (\bibinfo{year}{2024}).
\newblock


\bibitem[Dubois et~al\mbox{.}(2024)]%
        {10}
\bibfield{author}{\bibinfo{person}{Yann Dubois}, \bibinfo{person}{Chen~Xuechen Li}, \bibinfo{person}{Rohan Taori}, \bibinfo{person}{Tianyi Zhang}, \bibinfo{person}{Ishaan Gulrajani}, \bibinfo{person}{Jimmy Ba}, \bibinfo{person}{Carlos Guestrin}, \bibinfo{person}{Percy~S Liang}, {and} \bibinfo{person}{Tatsunori~B Hashimoto}.} \bibinfo{year}{2024}\natexlab{}.
\newblock \showarticletitle{Alpacafarm: A simulation framework for methods that learn from human feedback}.
\newblock \bibinfo{journal}{\emph{Advances in Neural Information Processing Systems}}  \bibinfo{volume}{36} (\bibinfo{year}{2024}).
\newblock


\bibitem[Fradkin and Holtz(2023)]%
        {if1}
\bibfield{author}{\bibinfo{person}{Andrey Fradkin} {and} \bibinfo{person}{David Holtz}.} \bibinfo{year}{2023}\natexlab{}.
\newblock \showarticletitle{Do incentives to review help the market? Evidence from a field experiment on Airbnb}.
\newblock \bibinfo{journal}{\emph{Marketing Science}} \bibinfo{volume}{42}, \bibinfo{number}{5} (\bibinfo{year}{2023}), \bibinfo{pages}{853--865}.
\newblock


\bibitem[Gao et~al\mbox{.}(2023a)]%
        {11}
\bibfield{author}{\bibinfo{person}{Chen Gao}, \bibinfo{person}{Xiaochong Lan}, \bibinfo{person}{Zhihong Lu}, \bibinfo{person}{Jinzhu Mao}, \bibinfo{person}{Jinghua Piao}, \bibinfo{person}{Huandong Wang}, \bibinfo{person}{Depeng Jin}, {and} \bibinfo{person}{Yong Li}.} \bibinfo{year}{2023}\natexlab{a}.
\newblock \showarticletitle{S3: Social-network Simulation System with Large Language Model-Empowered Agents}.
\newblock \bibinfo{journal}{\emph{arXiv preprint arXiv:2307.14984}} (\bibinfo{year}{2023}).
\newblock


\bibitem[Gao et~al\mbox{.}(2023b)]%
        {tempo4}
\bibfield{author}{\bibinfo{person}{Yiming Gao}, \bibinfo{person}{Feiyu Liu}, \bibinfo{person}{Liang Wang}, \bibinfo{person}{Zhenjie Lian}, \bibinfo{person}{Weixuan Wang}, \bibinfo{person}{Siqin Li}, \bibinfo{person}{Xianliang Wang}, \bibinfo{person}{Xianhan Zeng}, \bibinfo{person}{Rundong Wang}, \bibinfo{person}{Jiawei Wang}, {et~al\mbox{.}}} \bibinfo{year}{2023}\natexlab{b}.
\newblock \showarticletitle{Towards effective and interpretable human-agent collaboration in moba games: A communication perspective}.
\newblock \bibinfo{journal}{\emph{arXiv preprint arXiv:2304.11632}} (\bibinfo{year}{2023}).
\newblock


\bibitem[Hadi et~al\mbox{.}(2023)]%
        {bias1}
\bibfield{author}{\bibinfo{person}{Muhammad~Usman Hadi}, \bibinfo{person}{Rizwan Qureshi}, \bibinfo{person}{Abbas Shah}, \bibinfo{person}{Muhammad Irfan}, \bibinfo{person}{Anas Zafar}, \bibinfo{person}{Muhammad~Bilal Shaikh}, \bibinfo{person}{Naveed Akhtar}, \bibinfo{person}{Jia Wu}, \bibinfo{person}{Seyedali Mirjalili}, {et~al\mbox{.}}} \bibinfo{year}{2023}\natexlab{}.
\newblock \showarticletitle{A survey on large language models: Applications, challenges, limitations, and practical usage}.
\newblock \bibinfo{journal}{\emph{Authorea Preprints}}  \bibinfo{volume}{3} (\bibinfo{year}{2023}).
\newblock


\bibitem[Harris et~al\mbox{.}(2022)]%
        {27}
\bibfield{author}{\bibinfo{person}{Camille Harris}, \bibinfo{person}{Matan Halevy}, \bibinfo{person}{Ayanna Howard}, \bibinfo{person}{Amy Bruckman}, {and} \bibinfo{person}{Diyi Yang}.} \bibinfo{year}{2022}\natexlab{}.
\newblock \showarticletitle{Exploring the role of grammar and word choice in bias toward african american english (aae) in hate speech classification}. In \bibinfo{booktitle}{\emph{Proceedings of the 2022 ACM Conference on Fairness, Accountability, and Transparency}}. \bibinfo{pages}{789--798}.
\newblock


\bibitem[Hewitt et~al\mbox{.}(2024)]%
        {auto2}
\bibfield{author}{\bibinfo{person}{Luke Hewitt}, \bibinfo{person}{Ashwini Ashokkumar}, \bibinfo{person}{Isaias Ghezae}, {and} \bibinfo{person}{Robb Willer}.} \bibinfo{year}{2024}\natexlab{}.
\newblock \showarticletitle{Predicting results of social science experiments using large language models}.
\newblock \bibinfo{journal}{\emph{Preprint}} (\bibinfo{year}{2024}).
\newblock


\bibitem[Hoogendoorn et~al\mbox{.}(2017)]%
        {if3}
\bibfield{author}{\bibinfo{person}{Sander Hoogendoorn}, \bibinfo{person}{Simon~C Parker}, {and} \bibinfo{person}{Mirjam Van~Praag}.} \bibinfo{year}{2017}\natexlab{}.
\newblock \showarticletitle{Smart or diverse start-up teams? Evidence from a field experiment}.
\newblock \bibinfo{journal}{\emph{Organization Science}} \bibinfo{volume}{28}, \bibinfo{number}{6} (\bibinfo{year}{2017}), \bibinfo{pages}{1010--1028}.
\newblock


\bibitem[Horton(2023)]%
        {12}
\bibfield{author}{\bibinfo{person}{John~J Horton}.} \bibinfo{year}{2023}\natexlab{}.
\newblock \bibinfo{booktitle}{\emph{Large language models as simulated economic agents: What can we learn from homo silicus?}}
\newblock \bibinfo{type}{{T}echnical {R}eport}. \bibinfo{institution}{National Bureau of Economic Research}.
\newblock


\bibitem[Khorramrouz et~al\mbox{.}(2023)]%
        {ijk7}
\bibfield{author}{\bibinfo{person}{Adel Khorramrouz}, \bibinfo{person}{Sujan Dutta}, {and} \bibinfo{person}{Ashiqur~R KhudaBukhsh}.} \bibinfo{year}{2023}\natexlab{}.
\newblock \showarticletitle{For women, life, freedom: a participatory AI-based social web analysis of a watershed moment in Iran's gender struggles}.
\newblock \bibinfo{journal}{\emph{arXiv preprint arXiv:2307.03764}} (\bibinfo{year}{2023}).
\newblock


\bibitem[Leng and Yuan(2023)]%
        {15}
\bibfield{author}{\bibinfo{person}{Yan Leng} {and} \bibinfo{person}{Yuan Yuan}.} \bibinfo{year}{2023}\natexlab{}.
\newblock \showarticletitle{Do LLM Agents Exhibit Social Behavior?}
\newblock \bibinfo{journal}{\emph{arXiv preprint arXiv:2312.15198}} (\bibinfo{year}{2023}).
\newblock


\bibitem[Levitt and List(2009)]%
        {temp1}
\bibfield{author}{\bibinfo{person}{Steven~D Levitt} {and} \bibinfo{person}{John~A List}.} \bibinfo{year}{2009}\natexlab{}.
\newblock \showarticletitle{Field experiments in economics: The past, the present, and the future}.
\newblock \bibinfo{journal}{\emph{European Economic Review}} \bibinfo{volume}{53}, \bibinfo{number}{1} (\bibinfo{year}{2009}), \bibinfo{pages}{1--18}.
\newblock


\bibitem[Li et~al\mbox{.}(2020)]%
        {if2}
\bibfield{author}{\bibinfo{person}{Jia Li}, \bibinfo{person}{Noah Lim}, {and} \bibinfo{person}{Hua Chen}.} \bibinfo{year}{2020}\natexlab{}.
\newblock \showarticletitle{Examining salesperson effort allocation in teams: A randomized field experiment}.
\newblock \bibinfo{journal}{\emph{Marketing Science}} \bibinfo{volume}{39}, \bibinfo{number}{6} (\bibinfo{year}{2020}), \bibinfo{pages}{1122--1141}.
\newblock


\bibitem[Lin et~al\mbox{.}(2024)]%
        {bias4}
\bibfield{author}{\bibinfo{person}{Luyang Lin}, \bibinfo{person}{Lingzhi Wang}, \bibinfo{person}{Jinsong Guo}, {and} \bibinfo{person}{Kam-Fai Wong}.} \bibinfo{year}{2024}\natexlab{}.
\newblock \showarticletitle{Investigating bias in llm-based bias detection: Disparities between llms and human perception}.
\newblock \bibinfo{journal}{\emph{arXiv preprint arXiv:2403.14896}} (\bibinfo{year}{2024}).
\newblock


\bibitem[Liu et~al\mbox{.}(2023)]%
        {16}
\bibfield{author}{\bibinfo{person}{Pengfei Liu}, \bibinfo{person}{Weizhe Yuan}, \bibinfo{person}{Jinlan Fu}, \bibinfo{person}{Zhengbao Jiang}, \bibinfo{person}{Hiroaki Hayashi}, {and} \bibinfo{person}{Graham Neubig}.} \bibinfo{year}{2023}\natexlab{}.
\newblock \showarticletitle{Pre-train, prompt, and predict: A systematic survey of prompting methods in natural language processing}.
\newblock \bibinfo{journal}{\emph{Comput. Surveys}} \bibinfo{volume}{55}, \bibinfo{number}{9} (\bibinfo{year}{2023}), \bibinfo{pages}{1--35}.
\newblock


\bibitem[Luciano et~al\mbox{.}(2025)]%
        {if4}
\bibfield{author}{\bibinfo{person}{Margaret~M Luciano}, \bibinfo{person}{Jean~B Leslie}, \bibinfo{person}{John~E Mathieu}, \bibinfo{person}{Emily~R Hoole}, \bibinfo{person}{Rebecca Anderson}, {and} \bibinfo{person}{Virgil~W Fenters}.} \bibinfo{year}{2025}\natexlab{}.
\newblock \showarticletitle{Improving Virtual Team Collaboration Paradox Management: A Field Experiment}.
\newblock \bibinfo{journal}{\emph{Organization Science}} \bibinfo{volume}{36}, \bibinfo{number}{1} (\bibinfo{year}{2025}), \bibinfo{pages}{429--451}.
\newblock


\bibitem[Lucy and Bamman(2021)]%
        {bias5}
\bibfield{author}{\bibinfo{person}{Li Lucy} {and} \bibinfo{person}{David Bamman}.} \bibinfo{year}{2021}\natexlab{}.
\newblock \showarticletitle{Gender and representation bias in GPT-3 generated stories}. In \bibinfo{booktitle}{\emph{Proceedings of the third workshop on narrative understanding}}. \bibinfo{pages}{48--55}.
\newblock


\bibitem[Luo et~al\mbox{.}(2024)]%
        {17}
\bibfield{author}{\bibinfo{person}{Xiaoliang Luo}, \bibinfo{person}{Akilles Rechardt}, \bibinfo{person}{Guangzhi Sun}, \bibinfo{person}{Kevin~K Nejad}, \bibinfo{person}{Felipe Y{\'a}{\~n}ez}, \bibinfo{person}{Bati Yilmaz}, \bibinfo{person}{Kangjoo Lee}, \bibinfo{person}{Alexandra~O Cohen}, \bibinfo{person}{Valentina Borghesani}, \bibinfo{person}{Anton Pashkov}, {et~al\mbox{.}}} \bibinfo{year}{2024}\natexlab{}.
\newblock \showarticletitle{Large language models surpass human experts in predicting neuroscience results}.
\newblock \bibinfo{journal}{\emph{Nature human behaviour}} (\bibinfo{year}{2024}), \bibinfo{pages}{1--11}.
\newblock


\bibitem[Manning et~al\mbox{.}(2024)]%
        {18}
\bibfield{author}{\bibinfo{person}{Benjamin~S Manning}, \bibinfo{person}{Kehang Zhu}, {and} \bibinfo{person}{John~J Horton}.} \bibinfo{year}{2024}\natexlab{}.
\newblock \showarticletitle{Automated Social Science: A Structural Causal Model-Based Approach}.
\newblock  (\bibinfo{year}{2024}).
\newblock


\bibitem[Minson et~al\mbox{.}(2023)]%
        {tempo1}
\bibfield{author}{\bibinfo{person}{Julia~A Minson}, \bibinfo{person}{Corinne Bendersky}, \bibinfo{person}{Carsten de Dreu}, \bibinfo{person}{Eran Halperin}, {and} \bibinfo{person}{Juliana Schroeder}.} \bibinfo{year}{2023}\natexlab{}.
\newblock \showarticletitle{Experimental studies of conflict: Challenges, solutions, and advice to junior scholars}.
\newblock \bibinfo{journal}{\emph{Organizational Behavior and Human Decision Processes}}  \bibinfo{volume}{177} (\bibinfo{year}{2023}), \bibinfo{pages}{104257}.
\newblock


\bibitem[Namikoshi et~al\mbox{.}(2024)]%
        {19}
\bibfield{author}{\bibinfo{person}{Keiichi Namikoshi}, \bibinfo{person}{Alex Filipowicz}, \bibinfo{person}{David~A Shamma}, \bibinfo{person}{Rumen Iliev}, \bibinfo{person}{Candice~L Hogan}, {and} \bibinfo{person}{Nikos Arechiga}.} \bibinfo{year}{2024}\natexlab{}.
\newblock \showarticletitle{Using LLMs to Model the Beliefs and Preferences of Targeted Populations}.
\newblock \bibinfo{journal}{\emph{arXiv preprint arXiv:2403.20252}} (\bibinfo{year}{2024}).
\newblock


\bibitem[Oketunji et~al\mbox{.}(2023)]%
        {sentiment1}
\bibfield{author}{\bibinfo{person}{Abiodun~Finbarrs Oketunji}, \bibinfo{person}{Muhammad Anas}, {and} \bibinfo{person}{Deepthi Saina}.} \bibinfo{year}{2023}\natexlab{}.
\newblock \showarticletitle{Large Language Model (LLM) Bias Index--LLMBI}.
\newblock \bibinfo{journal}{\emph{arXiv preprint arXiv:2312.14769}} (\bibinfo{year}{2023}).
\newblock


\bibitem[Ray(2023)]%
        {20}
\bibfield{author}{\bibinfo{person}{Partha~Pratim Ray}.} \bibinfo{year}{2023}\natexlab{}.
\newblock \showarticletitle{ChatGPT: A comprehensive review on background, applications, key challenges, bias, ethics, limitations and future scope}.
\newblock \bibinfo{journal}{\emph{Internet of Things and Cyber-Physical Systems}}  \bibinfo{volume}{3} (\bibinfo{year}{2023}), \bibinfo{pages}{121--154}.
\newblock


\bibitem[Ren et~al\mbox{.}(2024)]%
        {ijk6}
\bibfield{author}{\bibinfo{person}{Siyue Ren}, \bibinfo{person}{Zhiyao Cui}, \bibinfo{person}{Ruiqi Song}, \bibinfo{person}{Zhen Wang}, {and} \bibinfo{person}{Shuyue Hu}.} \bibinfo{year}{2024}\natexlab{}.
\newblock \showarticletitle{Emergence of social norms in generative agent societies: principles and architecture}. In \bibinfo{booktitle}{\emph{Proceedings of the 33rd International Joint Conference on Artificial Intelligence (IJCAI)}}.
\newblock


\bibitem[Shanahan(2024)]%
        {22}
\bibfield{author}{\bibinfo{person}{Murray Shanahan}.} \bibinfo{year}{2024}\natexlab{}.
\newblock \showarticletitle{Talking about large language models}.
\newblock \bibinfo{journal}{\emph{Commun. ACM}} \bibinfo{volume}{67}, \bibinfo{number}{2} (\bibinfo{year}{2024}), \bibinfo{pages}{68--79}.
\newblock


\bibitem[Sun et~al\mbox{.}(2023)]%
        {gender1}
\bibfield{author}{\bibinfo{person}{Huaman Sun}, \bibinfo{person}{Jiaxin Pei}, \bibinfo{person}{Minje Choi}, {and} \bibinfo{person}{David Jurgens}.} \bibinfo{year}{2023}\natexlab{}.
\newblock \showarticletitle{Aligning with whom? large language models have gender and racial biases in subjective nlp tasks}.
\newblock \bibinfo{journal}{\emph{arXiv preprint arXiv:2311.09730}} (\bibinfo{year}{2023}).
\newblock


\bibitem[Tan et~al\mbox{.}(2024)]%
        {temp2}
\bibfield{author}{\bibinfo{person}{Zhen Tan}, \bibinfo{person}{Dawei Li}, \bibinfo{person}{Song Wang}, \bibinfo{person}{Alimohammad Beigi}, \bibinfo{person}{Bohan Jiang}, \bibinfo{person}{Amrita Bhattacharjee}, \bibinfo{person}{Mansooreh Karami}, \bibinfo{person}{Jundong Li}, \bibinfo{person}{Lu Cheng}, {and} \bibinfo{person}{Huan Liu}.} \bibinfo{year}{2024}\natexlab{}.
\newblock \showarticletitle{Large language models for data annotation: A survey}.
\newblock \bibinfo{journal}{\emph{arXiv preprint arXiv:2402.13446}} (\bibinfo{year}{2024}).
\newblock


\bibitem[Taubenfeld et~al\mbox{.}(2024a)]%
        {23}
\bibfield{author}{\bibinfo{person}{Amir Taubenfeld}, \bibinfo{person}{Yaniv Dover}, \bibinfo{person}{Roi Reichart}, {and} \bibinfo{person}{Ariel Goldstein}.} \bibinfo{year}{2024}\natexlab{a}.
\newblock \showarticletitle{Systematic biases in LLM simulations of debates}.
\newblock \bibinfo{journal}{\emph{arXiv preprint arXiv:2402.04049}} (\bibinfo{year}{2024}).
\newblock


\bibitem[Taubenfeld et~al\mbox{.}(2024b)]%
        {bias3}
\bibfield{author}{\bibinfo{person}{Amir Taubenfeld}, \bibinfo{person}{Yaniv Dover}, \bibinfo{person}{Roi Reichart}, {and} \bibinfo{person}{Ariel Goldstein}.} \bibinfo{year}{2024}\natexlab{b}.
\newblock \showarticletitle{Systematic biases in LLM simulations of debates}.
\newblock \bibinfo{journal}{\emph{arXiv preprint arXiv:2402.04049}} (\bibinfo{year}{2024}).
\newblock


\bibitem[Van~Heeswijk et~al\mbox{.}(2020)]%
        {if7}
\bibfield{author}{\bibinfo{person}{Wouter~JA Van~Heeswijk}, \bibinfo{person}{Martijn~RK Mes}, \bibinfo{person}{JMJ Schutten}, {and} \bibinfo{person}{WHM Zijm}.} \bibinfo{year}{2020}\natexlab{}.
\newblock \showarticletitle{Evaluating urban logistics schemes using agent-based simulation}.
\newblock \bibinfo{journal}{\emph{Transportation science}} \bibinfo{volume}{54}, \bibinfo{number}{3} (\bibinfo{year}{2020}), \bibinfo{pages}{651--675}.
\newblock


\bibitem[Wang et~al\mbox{.}(2023)]%
        {ijk4}
\bibfield{author}{\bibinfo{person}{Xinru Wang}, \bibinfo{person}{Chen Liang}, {and} \bibinfo{person}{Ming Yin}.} \bibinfo{year}{2023}\natexlab{}.
\newblock \showarticletitle{The Effects of AI Biases and Explanations on Human Decision Fairness: A Case Study of Bidding in Rental Housing Markets.}. In \bibinfo{booktitle}{\emph{IJCAI}}. \bibinfo{pages}{3076--3084}.
\newblock


\bibitem[Wei et~al\mbox{.}(2022)]%
        {24}
\bibfield{author}{\bibinfo{person}{Jason Wei}, \bibinfo{person}{Xuezhi Wang}, \bibinfo{person}{Dale Schuurmans}, \bibinfo{person}{Maarten Bosma}, \bibinfo{person}{Fei Xia}, \bibinfo{person}{Ed Chi}, \bibinfo{person}{Quoc~V Le}, \bibinfo{person}{Denny Zhou}, {et~al\mbox{.}}} \bibinfo{year}{2022}\natexlab{}.
\newblock \showarticletitle{Chain-of-thought prompting elicits reasoning in large language models}.
\newblock \bibinfo{journal}{\emph{Advances in neural information processing systems}}  \bibinfo{volume}{35} (\bibinfo{year}{2022}), \bibinfo{pages}{24824--24837}.
\newblock


\bibitem[Xu et~al\mbox{.}(2024)]%
        {tempo3}
\bibfield{author}{\bibinfo{person}{Ruoxi Xu}, \bibinfo{person}{Yingfei Sun}, \bibinfo{person}{Mengjie Ren}, \bibinfo{person}{Shiguang Guo}, \bibinfo{person}{Ruotong Pan}, \bibinfo{person}{Hongyu Lin}, \bibinfo{person}{Le Sun}, {and} \bibinfo{person}{Xianpei Han}.} \bibinfo{year}{2024}\natexlab{}.
\newblock \showarticletitle{AI for social science and social science of AI: A survey}.
\newblock \bibinfo{journal}{\emph{Information Processing \& Management}} \bibinfo{volume}{61}, \bibinfo{number}{3} (\bibinfo{year}{2024}), \bibinfo{pages}{103665}.
\newblock


\bibitem[Zhang et~al\mbox{.}(2020)]%
        {if8}
\bibfield{author}{\bibinfo{person}{Jingjing Zhang}, \bibinfo{person}{Gediminas Adomavicius}, \bibinfo{person}{Alok Gupta}, {and} \bibinfo{person}{Wolfgang Ketter}.} \bibinfo{year}{2020}\natexlab{}.
\newblock \showarticletitle{Consumption and performance: Understanding longitudinal dynamics of recommender systems via an agent-based simulation framework}.
\newblock \bibinfo{journal}{\emph{Information Systems Research}} \bibinfo{volume}{31}, \bibinfo{number}{1} (\bibinfo{year}{2020}), \bibinfo{pages}{76--101}.
\newblock


\bibitem[Zhang and Gao(2023)]%
        {26}
\bibfield{author}{\bibinfo{person}{Xuan Zhang} {and} \bibinfo{person}{Wei Gao}.} \bibinfo{year}{2023}\natexlab{}.
\newblock \showarticletitle{Towards llm-based fact verification on news claims with a hierarchical step-by-step prompting method}.
\newblock \bibinfo{journal}{\emph{arXiv preprint arXiv:2310.00305}} (\bibinfo{year}{2023}).
\newblock


\end{thebibliography}

\clearpage
\appendix

\section{Data}

\label{app:data}
\begin{figure}[H]
  \centering
  \includegraphics[width=\linewidth]{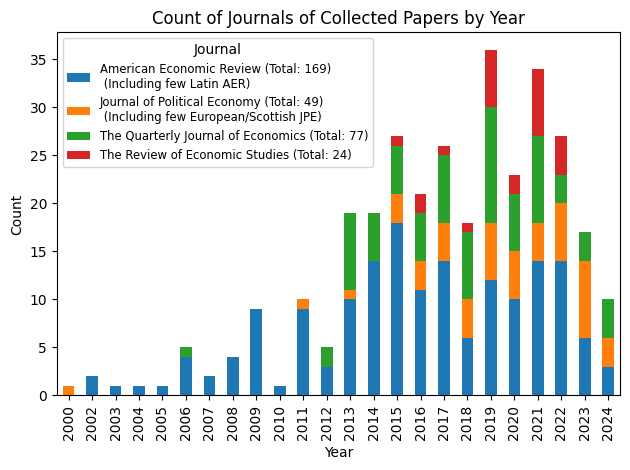}
  \caption{A Summary of Qualified Papers.}
  \Description{A bar chart showing the number of selected field experiment papers across different years or journals, highlighting the temporal or disciplinary distribution of the 276 papers used.}
  \label{fig:2}
\end{figure}

\section{Prompt Templates}

\subsection{The Prompt for Extractions}
\label{app:P_extract}
\begin{figure}[H]
  \centering
  \includegraphics[width=\linewidth]{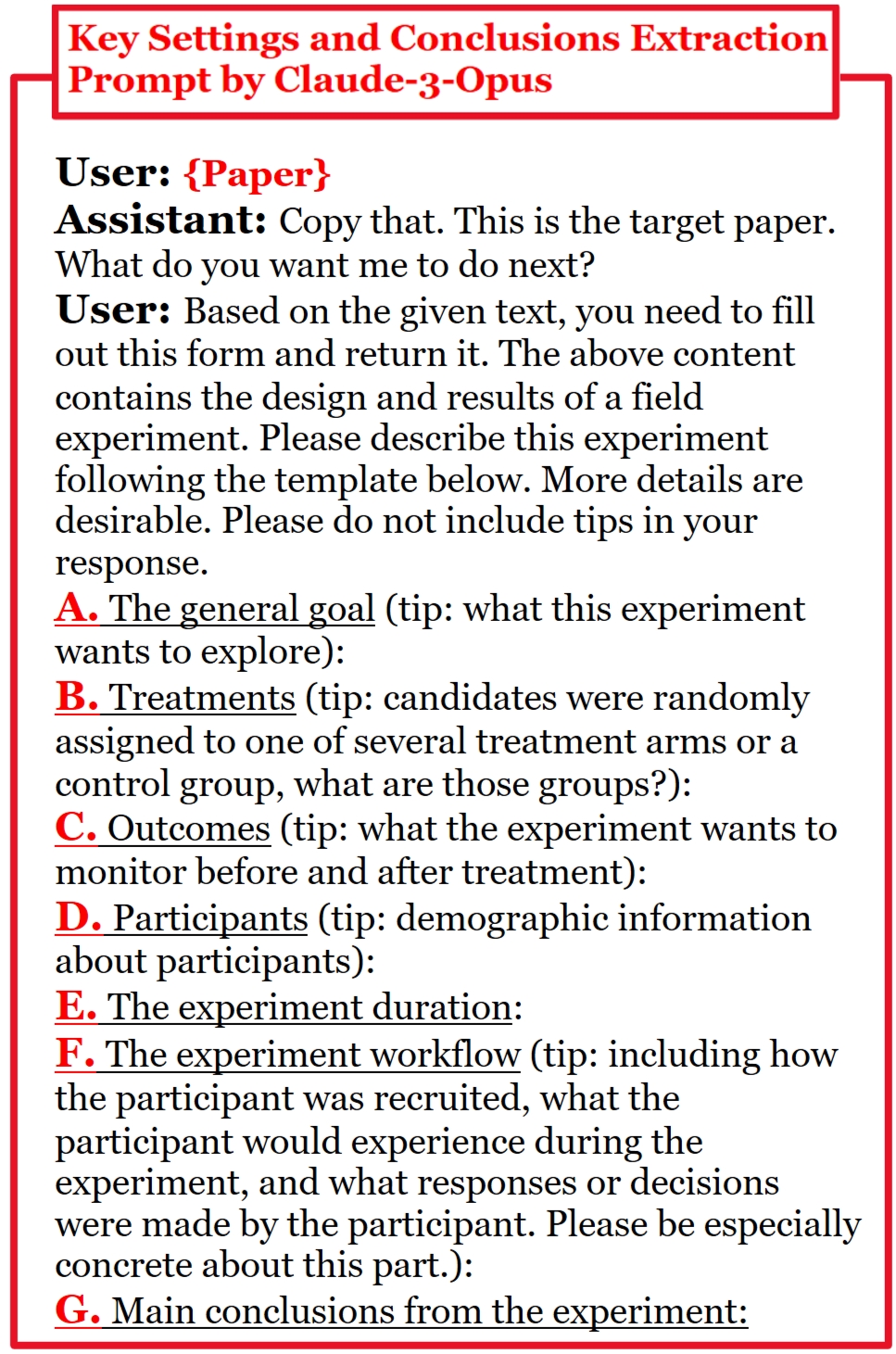}
  \caption{The Prompt for Extractions.}
  \Description{Screenshot of the prompt used to extract key field experiment settings and conclusions using Claude.}
  \label{fig:4}
\end{figure}

\subsection{The Prompt for Variant Generation}
\label{app:P_vg}
\begin{figure}[H]
  \centering
  \includegraphics[width=\linewidth]{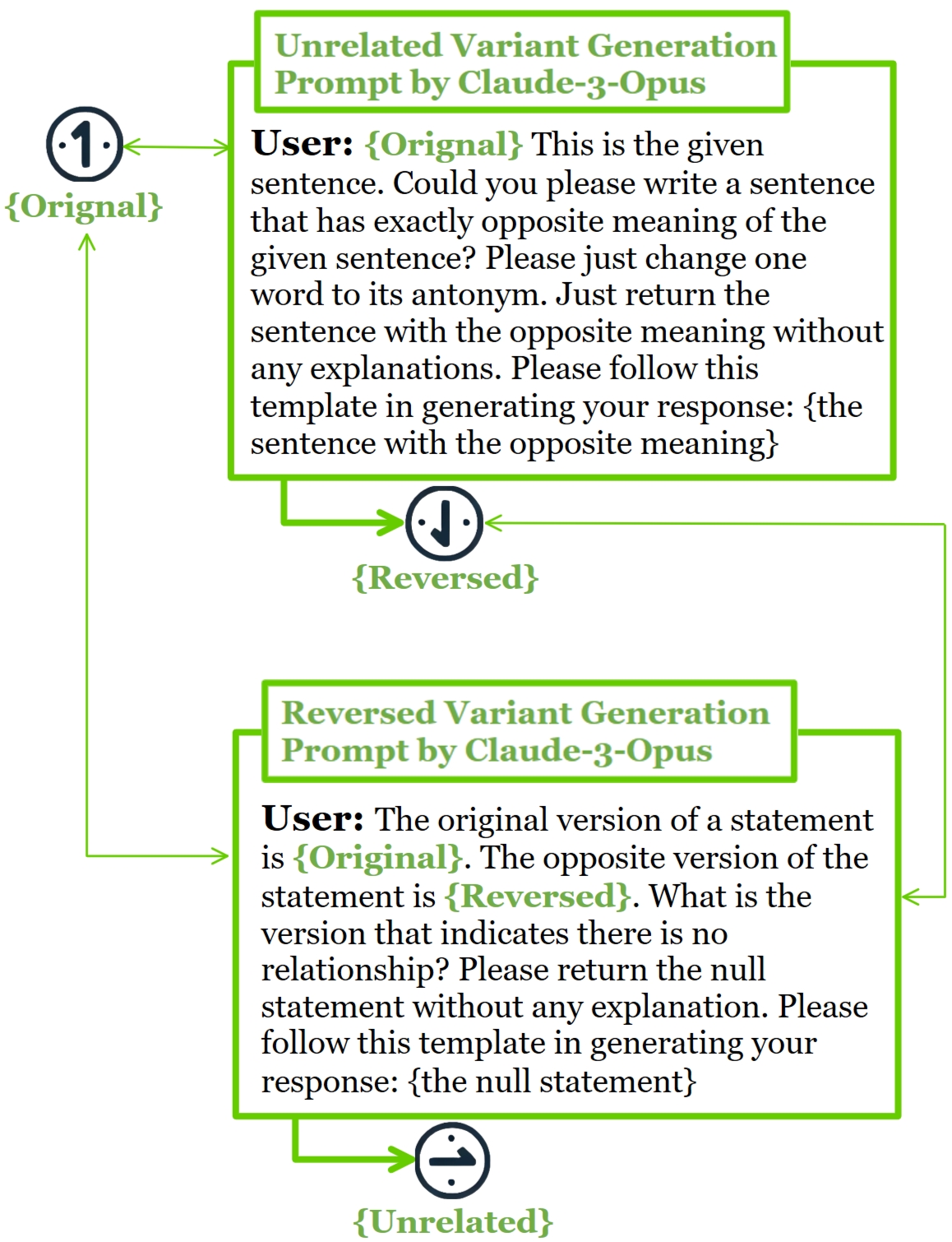}
  \caption{The Variant Generation Process.}
  \Description{Visual showing how reversed and unrelated variants are generated from the original experimental conclusion using LLM prompts.}

  \label{fig:5}
\end{figure}

Figure~\ref{fig:5} shows the extraction prompt used to retrieve experiment settings and conclusions from academic papers using Claude.

\subsection{Prompts for Prediction}
\label{app:P_pred}
\begin{figure}[H]
  \centering
  \includegraphics[width=\linewidth]{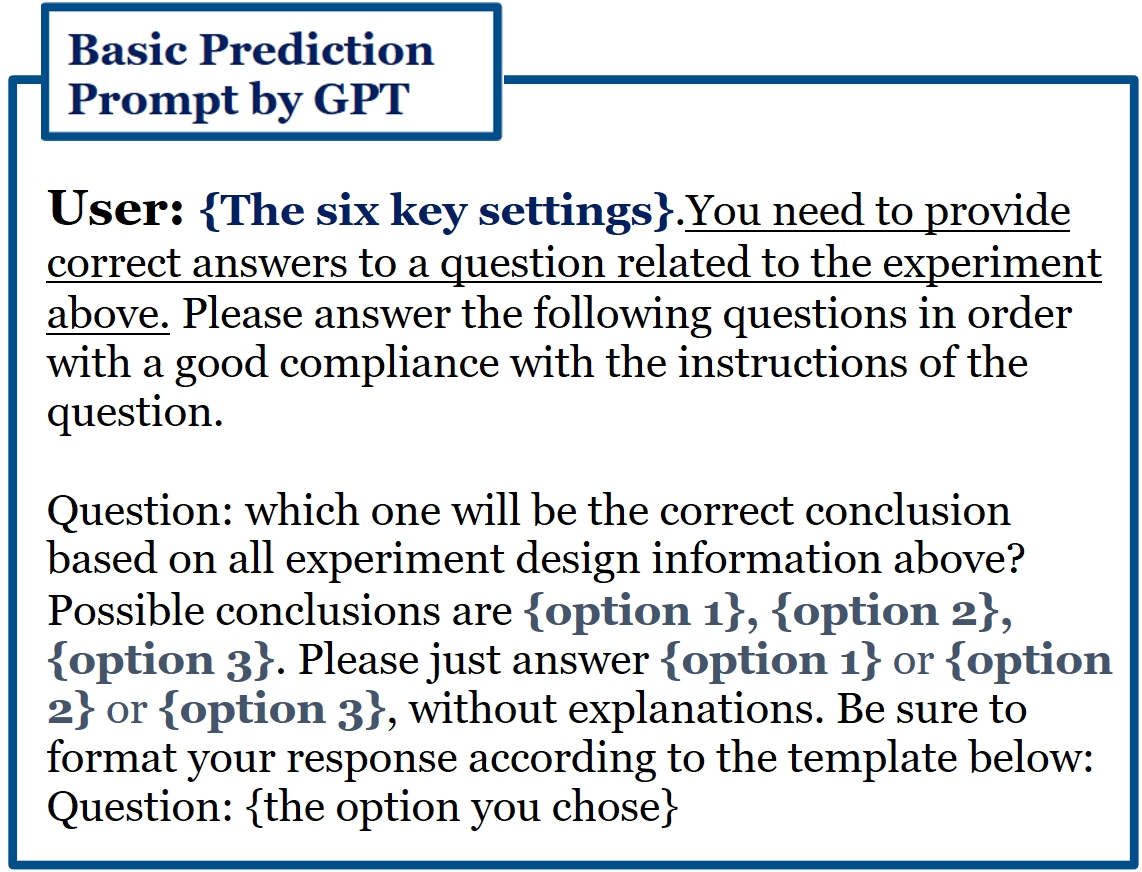}
  \caption{The Basic Prediction Prompt. }
  \Description{Prompt template used to perform basic prediction among multiple conclusions in a field experiment.}
  \label{fig:6}
\end{figure}

\begin{figure}[H]
  \centering
  \includegraphics[width=\linewidth]{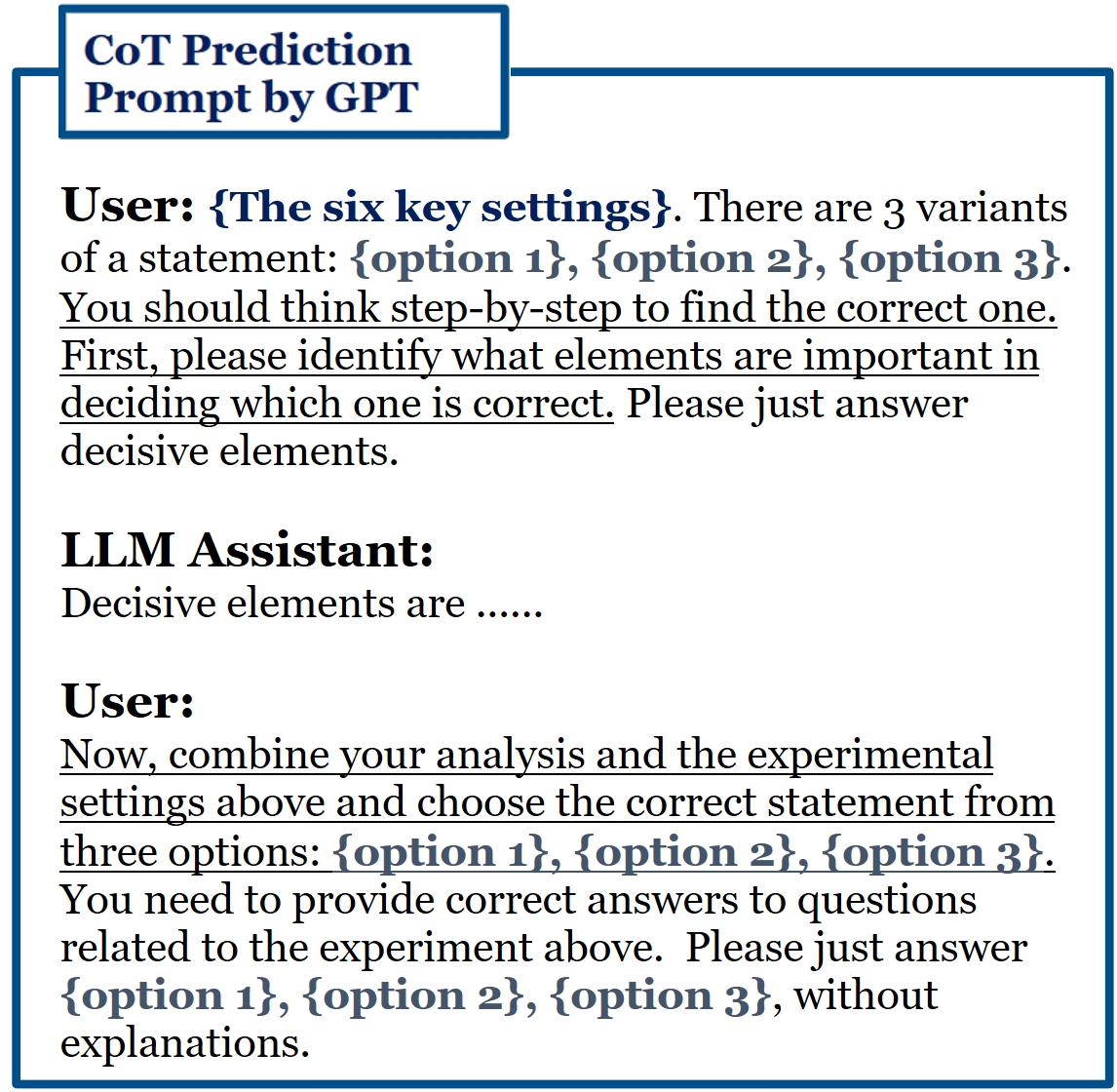}
  \caption{The Chain-of-Thought (CoT) Prediction Prompt. }
  \Description{Chain-of-thought prediction prompt where the model is instructed to reason through the decision before choosing a conclusion.}
  \label{fig:7}
\end{figure}

\subsection{Prompts for Topics and Sentiments Labeling}
\label{app:P_label}
\begin{figure}[H]
  \centering
  \includegraphics[width=\linewidth]{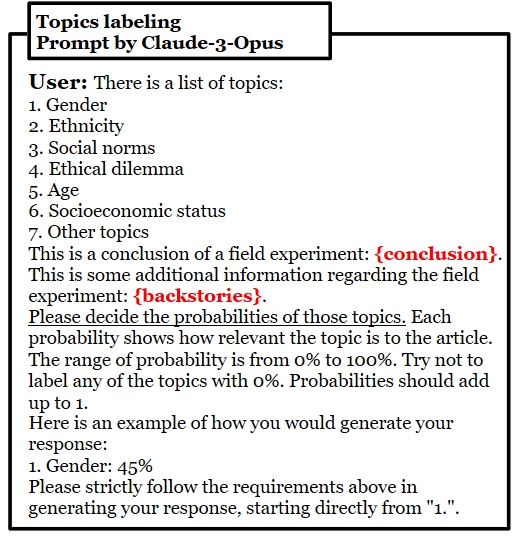}
  \caption{Topics Labeling Prompt.}
  \Description{Prompt used to assign topic relevance scores (e.g., gender, age, ethnicity) to each experimental context.}
  \label{fig:10a}
\end{figure}

\begin{figure}[H]
  \centering
  \includegraphics[width=\linewidth]{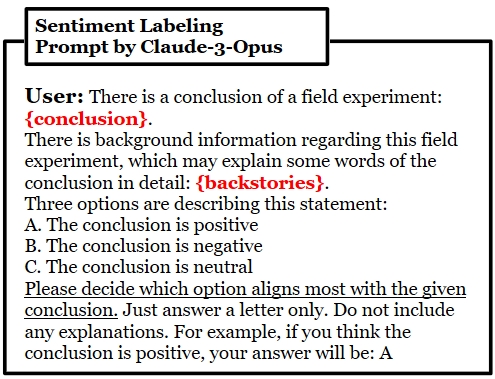}
  \caption{Sentiment Labeling Prompt.}
  \Description{Prompt used to classify the sentiment (positive, neutral, negative) associated with a conclusion.}
  \label{fig:10b}
\end{figure}

\begin{figure}[H]
  \centering
  \includegraphics[width=\linewidth]{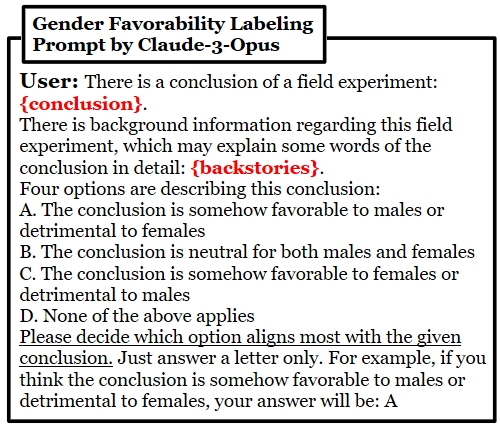}
  \caption{Gender Favorability Labeling Prompt.}
 \Description{Prompt used to label whether a conclusion's framing is favorable or unfavorable toward a specific gender.}
  \label{fig:10c}
\end{figure}

\section{Regression Variables}
\label{app:P_rv}
\begin{table}[H]
  \caption{Regression Variables}
  \label{tab:variables}
  \footnotesize
  \centering
  \begin{tabular}{p{0.28\linewidth} p{0.65\linewidth}}
    \toprule
    \textbf{Variable} & \textbf{Description} \\
    \midrule
    Conclusion Accuracy (DV) & Conclusion accuracy defined in Equation~\ref{eq:1}. \\
    Gender & Percentage indicating how strongly the context is associated with gender, obtained via labeling in Figure~\ref{fig:10a}. \\
    Ethnicity & Same as above, for ethnicity. \\
    Social Norms & Same as above, for social norms. \\
    Ethical Dilemmas & Same as above, for ethical dilemmas. \\
    Age & Same as above, for age-related context. \\
    Socioeconomic Status & Same as above, for socioeconomic status. \\
    Other Topics & Same as above, for all remaining topics. \\
    Sentiment & Sentiment score: 1 (positive), 0 (neutral), -1 (negative), labeled via Figure~\ref{fig:10b}. \\
    Gender Favorability & Gender-specific score: 1 (pro-female or anti-male), 0 (neutral), -1 (pro-male or anti-female), from Figure~\ref{fig:10c}. \\
    \bottomrule
  \end{tabular}
\end{table}

\end{document}